\RequirePackage{fix-cm}
\documentclass[twocolumn,epjc3]{svjour3}  
\smartqed  
\RequirePackage{graphicx}
\RequirePackage{amsmath}
\RequirePackage{multicol}
\usepackage{subcaption}
\usepackage{xcolor}

\journalname{Eur. Phys. J. C}
\begin{document}

\title{Some remarks on Hayward-AdS black hole surrounded by a fluid of strings
}


\author{F. F. Nascimento\thanksref{e1, addr1}
        \and
        V. B. Bezerra\thanksref{e2,addr1} 
        \and 
        J. M. Toledo\thanksref{e3,addr1}
        \and
        G.A. Marques\thanksref{e4,addr2}}

\thankstext{e1}{e-mail: fran.nice.fisica@gmail.com}
\thankstext{e2}{e-mail: valdir@fisica.ufpb.br}
\thankstext{e3}{e-mail: jefferson.m.toledo@gmail.com}
\thankstext{e4}{e-mail: gmarques@df.ufcg.edu.br}

\institute{Departamento de Física, 
  Universidade Federal da Paraíba, Caixa Postal 5008, 58059-900, João Pessoa, PB, Brazil \label{addr1}
  \and
  Unidade Acadêmica de Física, 
  Universidade Federal de Campina Grande, Caixa Postal 10071, 58429-900, Campina Grande, PB, Brazil \label{addr2}}

\date{Received: date / Accepted: date}

\date{Received: date / Accepted: date}

\maketitle

\begin{abstract}

We obtain a class of solutions corresponding to a generalization of the Hayward black hole by solving the Einstein equations coupled to a particular nonlinear electromagnetic field. The generalization is realized by considering, additionally, the presence of the cosmological constant and a source corresponding to an anisotropic fluid, namely, a fluid of strings, that surrounds the black hole. We show that the obtained class of solutions preserves or does not the regularity of the original Hayward black hole solution, depending on the values of the parameter $\beta$ which labels the different solutions. We discuss the characteristics of the solutions, from the point of view of the singularities of spacetime, by examining the behavior of the Kretschmann scalar as well as of the geodesics concerning their completeness. We analyze some aspects of thermodynamics, particularizing one of the solutions obtained, namely, for $\beta =-1/2$, in which case the regularity of the Hayward black hole is preserved.
Some thermodynamic quantities are obtained and analyzed, for example, pressure, heat capacity, and the critical points, and we show how these quantities change for different values of the parameter $q$ associated with the original Hayward solution, as well as with the parameter $b$ associated with the presence of the fluid of strings. The phase transitions are also analyzed by using the equation of state and the Gibbs free energy.

\end{abstract}

\section{Introduction}
\label{sec1}

Shortly after the General Theory of Relativity formulation, Schwarzschild obtained the simplest vacuum solution of Einstein equations, corresponding to a static and spherically symmetric black hole \cite{schwarzschild1916uber}. This metric was generalized by Reissner and Nordström to include electric charge \cite{reissner1916eigengravitation,nordstrom1918een} and, some years later, by Kerr and Newman by 
considering
the rotation \cite{kerr1963gravitational,newman1965metric}. All mentioned black hole solutions have  singularities, but this fact does not necessarily
means divergence of curvatures or of the metric.

The existence of singular solutions may represent a failure in the General Theory of Relativity, which can probably be solved in the framework of a quantum theory of gravity \cite{capozziello2011extended}. On the other hand, the study of non-singular (regular) spacetime metrics in the General Theory of Relativity has grown since the first one was proposed by Bardeen, by using phenomenological assumptions \cite{bardeen1968non}. In the past few years, some regular metrics were proposed in literature \cite{Dymnikova:1992ux,mars1996models,ayon1998regular,ayon1999new,dymnikova2003spherically,frolov2016notes,sajadi2017nonlinear}. The properties of regular black holes were also studied, as, for example, the black hole thermodynamics \cite{saleh2018thermodynamics,molina2021thermodynamics,paul2023more,singh2020thermodynamics}, geodesics \cite{abbas2014geodesic,zhou2012geodesic} and quasinormal modes \cite{fernando2012quasinormal,flachi2013quasinormal,lin2013quasinormal,perez2018region}.

Among
the regular spacetimes that can be found in literature, we can mention the Hayward metric \cite{hayward2006formation}, which represents a static and spherically symmetric black hole that behaves like a de Sitter spacetime at the origin ($r \rightarrow 0$) and is asymptotically flat for $r \rightarrow \infty$. The Hayward model defines the formation of the black hole from an initial vacuum region with finite density and pressure, which vanishes for large distances \cite{hayward2006formation,kumar2020hayward}. It can  also be understood as an exact model of a black hole in the General
Theory
of 
Relativity coupled with nonlinear electrodynamics. The Hayward black hole has also attracted a lot of interest and several investigations concerning its thermodynamics \cite{molina2021thermodynamics,pourhassan2016effects}, geodesics \cite{abbas2014geodesic,pedraza2021geodesics} and quasinormal modes \cite{lin2013quasinormal,roy2022generalized} have been performed in recentyears.

In the late 1970s, Letelier obtained a gauge-invariant model of a cloud of strings  \cite{letelier1979clouds}, motivated by the string theory, in which the fundamental components of nature are one-dimensional objects, rather than point particles. Later on, Letelier generalized the model by considering a fluid of strings, which now has a pressure associated with it \cite{letelier1981жидкость}. The solution corresponding to a static black hole immersed in this fluid of strings was obtained by Soleng \cite{soleng1995dark}, who showed that this fluid produces a correction in the Newton force law proportional to $1/r$. Thus, the fluid of strings could be used, in principle, as a model to explain the rotation curves of galaxies.

One of the most important turning points in cosmology in the past few years was the discovery of the universe's accelerated expansion through the observation of supernovae explosion \cite{riess1998observational,perlmutter1999measurements,riess1999bvri}. From this accelerated expansion, we can infer that on a large-scale structure, there is a repulsive energy that causes a negative pressure. Many mathematical models have been proposed to represent this energy and, among them, the cosmological constant \cite{copeland2006dynamics}. Initially introduced by Einstein in the General Theory of
Relativity field equations to allow the existence of a static universe, the cosmological constant had the role changed to generate the negative pressure that induces the universe's accelerated expansion \cite{copeland2006dynamics}. The cosmological constant has also been associated with a thermodynamic pressure in black hole systems, which leads to very interesting consequences \cite{caldarelli2000thermodynamics,dolan2011cosmological}.

Since the works of Bekenstein and Hawking in the 1970s \cite{bekenstein1973black,hawking1974black,hawking1976black}, the thermodynamics of black holes have been studied, mostly because it can represent a connection between gravity and quantum mechanics. In these studies, the presence of a cosmological constant has an important role, since, as already mentioned, it can be interpreted as a thermodynamic pressure with a conjugated thermodynamic volume associated with it \cite{caldarelli2000thermodynamics}. With this correspondence, it is possible to study the black hole system using various thermodynamic potentials and calculate the correspondent intensive variables, in addition to being possible to analyze the black hole stability and phase transitions.

In this paper, we obtain a class of solutions that correspond to a generalization of the Hayward black hole solution, which we are calling the Hayward-AdS black hole surrounded by a fluid of strings. In other words, the regular Hayward black hole is immersed in a fluid of strings and simultaneously, the presence of the cosmological constant is taken into account. In this scenario, we discuss the singularity of the spacetime, for different values of $\beta$. It is worth calling attention to the fact that the particular solution we have obtained, namely, $\beta=2$, was considered in the literature as a possible solution
that can mimic a perfect fluid dark matter\cite{soleng1995dark,Zhang:2020mxi}.
We  perform a detailed study of the black hole thermodynamics, analyzing the system stability and phase transitions.

This paper is organized as follows. In Sec. \ref{sec2}, we obtain a class of solutions corresponding to the Hayward-AdS black hole surrounded by a fluid of string, which generalizes the Hayward black hole solution in the sense that a cloud of strings was considered\cite{nascimento2024some} and discuss when the obtained solutions are regular of singular, by analyzing the appropriate behavior of the Kretschmann scalar and the geodesics with respect to their completeness
or incompleteness. In. Sec. \ref{sec:therm}, we analyze the thermodynamics of black holes, with emphasis to the behavior of some thermodynamic quantities in which concerns their dependence on the intensity of the fluid of strings. Finally, in Sec. \ref{sec:concl}, we present our conclusions.



\section{Hayward-AdS black hole surrounded by a fluid of strings.}
\label{sec2}

\subsection{Introduction}

Let us start by considering the Hayward black hole spacetime \cite{hayward2006formation,nascimento2024some} which can be understood as arising from the solution of Einstein's equation coupled to a nonlinear electromagnetic field. In this scenario, the physical source of this black hole is a nonlinear electromagnetic field \cite{molina2021thermodynamics,bronnikov2001regular,fan2016construction}.

In this section, we obtain a class of spherically symmetric solutions that generalizes the original Hayward one due to the presence of two other sources, namely, the cosmological constant and a fluid of strings.

The action that describes appropriately a system minimally coupled to a nonlinear electromagnetic field can be written as
\begin{equation}
S= \frac{1}{16\pi}\int d^4x\sqrt{-g}(R+\mathcal{L})\label{eq_action}
\end{equation}
where $g$ is the determinant of the metric tensor, $g_{\mu\nu}$, $R$ is the scalar curvature and $\mathcal{L}$ is the Lagrangian density of the nonlinear electromagnetic field \cite{bronnikov2001regular,fan2016construction,molina2021thermodynamics}.

From this coupling between gravitational and nonlinear electromagnetic fields, we get the following equations \cite{bronnikov2001regular}, by varying the action given by Eq.(\ref{eq_action}), concerning the metric
\begin{equation}
G_{\mu\nu} = 2 \left(\frac{\partial\mathcal{L}(F)}{\partial F}F_{\mu\sigma}F^{\sigma}_{\,\,\nu}-\frac{1}{4}g_{\mu\nu}\mathcal{L}(F)\right),
\label{equação de einstein mista}
\end{equation}
where $G_{\mu\nu}=R_{\mu\nu}-\frac{1}{2}Rg_{\mu\nu}$.

In Hayward's solution, the source comes from nonlinear electrodynamics where the Lagrangian density is written as
\begin{equation}
\mathcal{L}(F)=\frac{6 \left(2l^2F\right)^{3/2}}{\kappa ^2 l^2\left[1+\left(2l^2F\right)^{3/4}\right]^2}.
\label{Hayward's Lagrangian}
\end{equation}

\noindent which represents the particular source of nonlinear electrodynamics used to derive the Hayward black hole. In Eq. (\ref{Hayward's Lagrangian}), $l$ is Hayward’s parameter, whose value, in principle, will be restricted
to the range $0\leq l<\infty$, and the Lagrangian is a nonlinear function of the electromagnetic scalar $F=F^{\mu\nu}F_{\mu\nu}$, where $F_{\mu\nu}$ is the
Maxwell-Faraday tensor. For a spherically symmetric space-time that is only
magnetically charged, the only nonzero component of $F_{\mu\nu}$ is \cite{bronnikov2001regular}

\begin{equation}\label{F23}
F_{23}=q_m \sin{\theta},
\end{equation}

\noindent and the scalar $F$ is

\begin{equation}\label{F}
F=\frac{2q_m^2}{r^4}.
\end{equation}

In weak field limit, the Lagrangian of Eq. (\ref{Hayward's Lagrangian}) behaves like $l^{-1 F^{3/2}}$, which is different from Maxwell field. This is the expected behavior for Hayward blackholes \cite{fan2016construction,fan2017critical}, since regular center requires a non-Maxwell behavior of Lagrangian\cite{bronnikov2017comment}.

 It should be pointed out that, in his original work, Hayward had considered the parameter $l$ to be of the order of Planck’s length, and it was then related to a magnetic charge through the definition \cite{molina2021thermodynamics,bronnikov2001regular,fan2016construction,bronnikov2017comment,toshmatov2018comment}

\begin{eqnarray}\label{magnetic charge}
q_m=\frac{\sqrt[3]{r_s^2l}}{2}=\frac{\sqrt[3]{4m^2l}}{2}.
\end{eqnarray}

Thus, it is possible to obtain the following components for the Hayward stress-energy tensor \cite{hayward2006formation}:

\begin{equation}
T_t^{\;t}=T_r^{\;r}=\frac{12l^2m^2}{(r^3+2l^2m)^2},
\label{eq:1.6}
\end{equation}
\begin{equation}
T_\theta^{\;\theta}=T_\phi^{\;\phi}=-\frac{24(r^3-l^2m)l^2m^2}{(r^3+2l^2m)^3},
\label{eq:1.7}
\end{equation}
 
 \noindent where $(l,m)$ are constants that we will assume to be positive.

In order to consider the cosmological constant and the fluid of strings, we proceed by adding the term $-\Lambda g_{\mu\nu}$ to the left-hand side of Eq. (\ref{equação de einstein mista}) and the energy-momentum tensor corresponding to the fluid of strings to the right-hand side of the same equation, which results in
\begin{equation}
\begin{aligned}
&R_{\mu\nu}-\frac{1}{2}Rg_{\mu\nu}-\Lambda g_{\mu\nu}= 
\\
&2 \left(\frac{\partial\mathcal{L}(F)}{\partial F}F_{\mu\sigma}F^{\sigma}_{\,\,\nu}-\frac{1}{4}g_{\mu\nu}\mathcal{L}(F)\right)+ T_{\mu\nu}^{\text{FS}},
\label{geral_eq}
\end{aligned}
\end{equation}
where $T_{\mu\nu}^{\text{FS}}$ refers to the energy-momentum tensor of the fluid of strings. The right-hand side of the above equation is an effective energy-momentum tensor, with the first term associated with the modifications of Einstein equations due to the coupling with a nonlinear electromagnetic field, and the second refers to the fluid of strings.

%
%
\subsection{The energy-momentum tensor of the fluid of strings}

The world line of a moving particle with four-velocity given by $u^{\mu}=dx^\mu/d\lambda$, with $\lambda$ being an independent parameter, can be described by $x=x(\lambda)$. Otherwise, if we consider, instead of a particle, a moving infinitesimally thin string, the trajectory corresponds to a two-dimensional world sheet $\Sigma$, which can be obtained by \cite{letelier1979clouds}

\begin{equation}
x^{\mu}=x^{\mu}(\lambda^{a}), \; \;a=0,1,
\label{eq:1.10}
\end{equation}
with $\lambda_0$ and $\lambda_1$ being timelike and spacelike parameters, respectively. Therefore, instead of the
four-velocity, $u^{\mu}$, we have a bivector, $\Sigma^{\mu\nu}$, such that \cite{letelier1979clouds}

\begin{equation}
\Sigma^{\mu\nu}=\epsilon^{ab}\frac{\partial{x^\mu}}{\partial{\lambda^{a}}}\frac{\partial{x^\nu}}{\partial{\lambda^{b}}},
\label{eq:1.11}
\end{equation}
where $\epsilon^{ab}$ is the two-dimensional Levi-Civita symbol, with $\epsilon^{01}=-\epsilon^{10}=1$.

It is worth emphasizing that on this world sheet, there will be an induced metric, $\gamma_{ab}$, with
$a, b = 0, 1$, such that,

\begin{equation}
\gamma_{ab}=g_{\mu\nu}\frac{\partial{x^\mu}}{\partial{\lambda^{a}}}\frac{\partial{x^\nu}}{\partial{\lambda^{b}}},
\label{eq:1.12}
\end{equation}
whose determinant is given by $\gamma$.
The stress-energy tensor associated with a dust cloud is given by $T^{\mu\nu}=\rho u^{\mu}u^{\nu}$, with $u^{\mu}$ being the normalized four-velocity and $\rho$ being the proper density of the flow. Similarly, for a cloud of strings, we have \cite{letelier1979clouds}

\begin{equation}
T^{\mu\nu}=\rho\frac{\Sigma^{\mu\beta}\Sigma_{\beta}^{\;\nu}}{(-\gamma)^{1/2}},
\label{eq:1.13}
\end{equation}

\noindent where $\gamma=\frac{1}{2}\Sigma^{\mu\nu}\Sigma_{\mu\nu}$.

Now, if we take into account a perfect fluid with pressure $p$, it can be described by the stress-energy tensor $T^{\mu\nu}= (p+\rho)u^{\mu}u^{\nu}-p g^{\mu\nu}$. Similarly, considering a perfect fluid of strings with pressure $p$, we have the stress-energy tensor \cite{letelier1981жидкость}

\begin{equation}
T^{\mu\nu}=(p+\sqrt{-\gamma}\rho)\frac{\Sigma^{\mu\beta}\Sigma_\beta^{\;\nu}}{(-\gamma)}+p g^{\mu\nu}.
\label{eq:1.14}
\end{equation}

Considering the stress-energy tensor given by Eq. (\ref{eq:1.14}), Soleng obtained the metric corresponding to a spherically symmetrical static black hole surrounded by a fluid of strings \cite{soleng1995dark}. Then, we assume that the components of the stress-energy tensor are related through the equations

\begin{equation}
T_{t}^{\;t}= T_{r}^{\;r},
\label{eq:1.15}
\end{equation}
\begin{equation}
T_{t}^{\;t}= -\beta T_{\theta}^{\;\theta}= -\beta T_{\phi}^{\;\phi},
\label{eq:1.16}
\end{equation}
where $\beta$ is a dimensionless constant. 

The stress-energy tensor whose components are given by Eqs. (\ref{eq:1.15})-(\ref{eq:1.16}), was interpreted as being associated with a kind of anisotropic fluid with spherical symmetry \cite{dymnikova1992vacuum,soleng1994correction}. In these references, it was interpreted as the stress-energy tensor associated with anisotropic vacuum polarization in a spherically symmetric space-time.

It is worth calling attention to the fact that this stress-energy tensor has been used in different scenarios \cite{salgado2003simple,giambo2002anisotropic,dymnikova2002cosmological}. In particular, the conditions imposed by
Eq. (\ref{eq:1.16}) leads to a class of spherically symmetric solutions of Einstein’s field equations with two parameters \cite{salgado2003simple}.

In this article, we assume the components of the stress-energy tensor for the fluid of strings given by \cite{toledo2020black}:
\begin{equation}
T_{t}^{\;t}= T_{r}^{\;r}= -\frac{\epsilon}{r^2}\left(\frac{b}{r}\right)^{2/\beta},
\label{eq:1.17.1}
\end{equation}
\begin{equation}
T_{\theta}^{\;\theta}= T_{\phi}^{\;\phi}= \frac{\epsilon}{\beta r^2}\left(\frac{b}{r}\right)^{2/\beta},
\label{eq:1.18.1}
\end{equation}

\noindent where $b$ is a positive integration constant and $\epsilon=\pm1$ determines the sign of the energy density of the fluid of strings.
%
%
\subsection{Hayward black hole solution with cosmological constant and surrounded by a fluid of strings}

Now, let us consider Eq. (\ref{equação de einstein mista}) with the modifications introduced by the addition of the cosmological constant, as well as of the fluid of strings, whose components of the energy-momentum tensor for this fluid are given by Eq. (\ref{eq:1.17.1}) and Eq. (\ref{eq:1.18.1}). More precisely, we will consider the Eq. (\ref{geral_eq}).

The line element for an isotropic static and spherically symmetric spacetime, without loss of generality, can be written as follows \footnote{Here we use the natural units for $c = G = \hbar$}:
\begin{equation}
ds^2=e^\nu dt^2-e^\lambda dr^2-r^2 d\theta^2-r^2\sin^2\theta d\phi^2.
\label{eq:1.1}
\end{equation}

Thus, the Einstein field equations for the case under consideration, taking into account the presence of the cosmological constant and the additional source corresponding to the fluid of strings, can be written as

\begin{equation}
\begin{aligned}
&e^{-\lambda}\left(\frac{\lambda'}{r}-\frac{1}{r^2}\right)+\frac{1}{r^2}=
\\
&\frac{12l^2m^2}{(r^3+2l^2m)^2}+\Lambda-\frac{\epsilon}{r^2}\left(\frac{b}{r}\right)^{2/\beta},
\label{eq:1.17}
\end{aligned}
\end{equation}
\begin{equation}
\begin{aligned}
&-e^{-\lambda}\left(\frac{\nu'}{r}+\frac{1}{r^2}\right)+\frac{1}{r^2}=
\\
&\frac{12l^2m^2}{(r^3+2l^2m)^2}+\Lambda-\frac{\epsilon}{r^2}\left(\frac{b}{r}\right)^{2/\beta},
\label{eq:1.18}
\end{aligned}
\end{equation}
\begin{equation}
\begin{aligned}
&\frac{1}{2}e^{-\lambda}\left(\frac{\nu'\lambda'}{2}+\frac{\lambda'}{r}-\frac{\nu'}{r}-\frac{\nu'^2}{2}-\nu''\right)=
\\
&-\frac{24(r^3-l^2m)l^2m^2}{(r^3+2l^2m)^3}+\Lambda+\frac{\epsilon}{\beta r^2}\left(\frac{b}{r}\right)^{2/\beta}.
\label{eq:1.19}
\end{aligned}
\end{equation}

\noindent By subtracting Eqs. (\ref{eq:1.17}) and (\ref{eq:1.18}), we obtain that:

\begin{equation}
\lambda=-\nu\Rightarrow\lambda'=-\nu'.
\label{eq:1.20}
\end{equation}

\noindent Summing Eqs. (\ref{eq:1.17}) and (\ref{eq:1.18}) and considering eq. (\ref{eq:1.20}), after some simplifications, we obtain:

\begin{equation}
e^{-\lambda}\frac{\lambda'}{r}-e^{-\lambda}\frac{1}{r^2}+\frac{1}{r^2}=\frac{12l^2m^2}{(r^3+2l^2m)^2}+\Lambda-\frac{\epsilon}{r^2}\left(\frac{b}{r}\right)^{2/\beta}.
\label{eq:1.21}
\end{equation}

\noindent Now, let us write the following relations:

\begin{equation}
\nu=-\lambda=ln(1+f(r)).
\label{eq:1.22}
\end{equation}

\noindent Considering Eqs. (\ref{eq:1.20}) and (\ref{eq:1.22}), we can write Eqs. (\ref{eq:1.21}) and (\ref{eq:1.19}), respectively, as follows:

\begin{equation}
-\frac{1}{r^2}(rf'+f)=\frac{12l^2m^2}{(r^3+2l^2m)^2}+\Lambda-\frac{\epsilon}{r^2}\left(\frac{b}{r}\right)^{2/\beta},
\label{eq:1.25}
\end{equation}

\begin{equation}
2\frac{f'}{r}+f''=48\frac{(r^3-l^2m)l^2m^2}{(r^3+2l^2m)^3}-2\Lambda-\frac{2\epsilon}{\beta r^2}\left(\frac{b}{r}\right)^{2/\beta}.
\label{eq:1.26}
\end{equation}

\noindent Summing Eqs. (\ref{eq:1.25}) and (\ref{eq:1.26}) and multiplying the result by $r^2$, we get the following differential equation:

\begin{equation}
\begin{aligned}
&r^2f''+rf'-f+\Lambda r^2-\frac{12l^2m^2r^2}{(r^3+2l^2m)^2}
\\
&-48\frac{(r^3-l^2m)l^2m^2r^2}{(r^3+2l^2m)^3}+\epsilon\left(\frac{\beta+2}{\beta}\right)\left(\frac{b}{r}\right)^{2/\beta}=0.
\label{eq:1.27}
\end{aligned}
\end{equation}

\noindent whose solution is given by

\begin{equation}
\begin{aligned}
f(r)=&-\frac{2 m r^2}{r^3+2 l^2 m}-\frac{\Lambda  r^2}{3}+
\\
&\left\{\begin{array}{rcl}\epsilon b[1+2\log (r)]/2r\;\;$for$\;\;\beta=2,\\
\epsilon\beta(\beta -2)^{-1} \left(\frac{b}{r}\right)^{2/\beta}\;\;$for$\;\;\beta\neq 2.
\end{array}
\right.
\label{eq:1.28}
\end{aligned}
\end{equation}

\noindent Thus, substituting Eq. (\ref{eq:1.28}) into Eq. (\ref{eq:1.22}) and then into Eq. (\ref{eq:1.1}), we finally obtain the Hayward-AdS black hole surrounded by a fluid of strings :

\begin{equation}
ds^2=f(r) dt^2-f(r)^{-1} dr^2-r^2 d\theta^2-r^2\sin^2\theta d\phi^2,
\label{eq:line_element}
\end{equation}

\noindent where 

\begin{align}
f(r)&=1-\frac{2 m r^2}{r^3+2 l^2 m}-\frac{\Lambda  r^2}{3}+\nonumber
\\ 
&\bigg\{\begin{array}{rcl}\epsilon b[1+2\log (r)]/2r\,\,\text{for}\,\,\beta=2,\\
\epsilon\beta(\beta -2)^{-1} \left(\frac{b}{r}\right)^{2/\beta}\,\,\text{for}\,\,\beta\neq 2.
\end{array}
\label{eq:f_r}
\end{align}

We can recover some other solutions through this metric if we make the following choices displayed in the Table (\ref{table_dif_ST}).

\begin{table}[h]
\caption{Space-time that can be recovered by making $\Lambda=0$ in Eq. (\ref{eq:f_r}) and different values of $\beta$, $b$ and $\epsilon$.}
\begin{tabular}{lcclccclcccclcl}
\hline
&$\epsilon$ &     $b$             &     $\beta$                       &     space-time           &\\\hline
&$1$       &     ${R}^{*}$            &     $lim_{\beta\rightarrow\infty}$&     Hayward-Letelier             &\\
&$1$       &     $Q$             &     $1$                           &     Hayward-Reissner-Nordström &\\
&$1$       &     $\Lambda^{-1/2}$&     $-1$                          &     Hayward-de Sitter&\\\hline
\end{tabular}
\label{table_dif_ST}
\end{table}

It is worth calling attention to the fact that, formally, the class of solution obtained can be considered as generated by two fluids, with an effective energy-momentum tensor, one corresponding to a nonlinear electromagnetic source and the other to the fluid of strings, and with a cosmological constant. 

\subsection{Kretschmann scalar analysis}

Now, 
let us examine the existence of singularities of the metric
using the coordinate invariant quantities constructed with the curvature tensor, namely, the Kretschmann scalar given by $K=R_{\alpha\beta\mu\nu}R^{\alpha\beta\mu\nu}$.   
It is worth calling attention to the fact that the existence  of singularities in such curvature scalar does not imply that the spacetime is singular. Similarly, the absence of singularity,  or equivalently, the presence of regularity does not imply
that the spacetime is regular. At this point, it is important to call attention to the fact that the criterion to characterizespace-time is singular based on the behavior of the curvature scalar is just a preliminary one. In fact, 
the most widely accepted criterion to conclude if a given spacetime is singular  is through the incompleteness of geodesics\cite{wald2010general}. 
In fact, the divergence of the Kretschmann scalar does not imply that there is no complete geodesic, nor does geodesic incompleteness necessarily imply the existence of a physical singularity. 

Let us start by doing a preliminary analysis of the spacetime’s singularity or regularity of the metric obtained from the point of view of the scalar curvature invariant, specifically, to calculate the Kretschmann scalar
  for different values of the parameter $\beta$, and check what are the differences as compared with similar results for a Hayward black hole, which is regular everywhere. 
In the sequence, we will proceed with similar analysis, but from different point of view, namely, by  taking into account the geodesics in which concerns their completeness or incompleteness\cite{hawking2023large,wald2010general}. 

\subsubsection{For $\beta\neq2$}

In this case, the Kretschmann scalar is

\begin{equation}
\begin{aligned}
K=&R_{\alpha\beta\mu\nu}R^{\alpha\beta\mu\nu}=
\\
&+\frac{8 m^2 \left(r^3-4 l^2 m\right)^2}{\left(2 l^2 m+r^3\right)^4}+\frac{1024 l^6 m^5 \left(l^2 m-r^3\right)}{\left(2 l^2 m+r^3\right)^6}
\\
&+\frac{8 m^2 r^6 \left(444 l^4 m^2-44 l^2 m r^3+5 r^6\right)}{\left(2 l^2 m+r^3\right)^6}
\\
&+\frac{16 m \epsilon  \left(\frac{b}{r}\right)^{2/\beta } \left(4 l^2 m-r^3\right)}{(\beta -2) \left(2 l^2 m r+r^4\right)^2}+\frac{8 \epsilon ^2 \left(\frac{b}{r}\right)^{4/\beta }}{(\beta -2)^2 r^4}
\\
&-\frac{32 m \epsilon  \left(\frac{b}{r}\right)^{2/\beta } \left(4 l^4 m^2-14 l^2 m r^3+r^6\right)}{(\beta -2) \beta  r^2 \left(2 l^2 m+r^3\right)^3}
\\
&-\frac{16 m r \epsilon  \left(\frac{b}{r}\right)^{2/\beta } \left(4 (\beta -4) l^2 m+(\beta +2) r^3\right)}{(\beta -2) \left(2 l^2 m+r^3\right)^3}
\\
&+\frac{16 \Lambda ^2}{9}-\frac{64 (\beta -1) l^4 m^3 \epsilon  \left(\frac{b}{r}\right)^{2/\beta }}{(\beta -2) r^2 \left(2 l^2 m+r^3\right)^3}
\\
&+\frac{4 \left(\beta ^4+3 \beta ^2+4 \beta +4\right) \epsilon ^2 \left(\frac{b}{r}\right)^{4/\beta }}{(\beta -2)^2 \beta ^2 r^4}
\\
&+\frac{16 \Lambda  m \left(r^3-4 l^2 m\right)^2}{3 \left(2 l^2 m+r^3\right)^3}-\frac{8 (\beta -1) \Lambda  \epsilon  \left(\frac{b}{r}\right)^{2/\beta }}{3 \beta  r^2}
\\
&+\frac{8 \Lambda ^2}{9}+\frac{16 \Lambda  m \left(4 l^2 m-r^3\right)}{3 \left(2 l^2 m+r^3\right)^2}.
\label{eq:1.31}
\end{aligned}
\end{equation}

Next, we will determine the limit of the Kretschmann scalar when $r\rightarrow 0$ and $r\rightarrow \infty$ for $l>0$, $b>0$ and $\Lambda \in \mathbf{R}$ and $\epsilon=+1$. The analysis will only be done for some $\beta$ values.

\par(i) For $\beta<-1$, the Kretschmann scalar diverges very close to the origin and is finite in a region very far from the black hole. 
\begin{equation}
\lim_{r\rightarrow 0}K=\infty.
\end{equation} 
\begin{equation}
\lim_{r\rightarrow \infty}K=\frac{8 \Lambda ^2}{3}.
\end{equation}

\par(ii) For $\beta=-1$, the Kretschmann scalar is finite, and are given by
\begin{equation}
\lim_{r\rightarrow 0}K=\frac{8 \left(l^2 -b^2 \left(\Lambda  l^2+3\right)\right)^2}{3 b^4 l^4}.
\end{equation}
\begin{equation}
\lim_{r\rightarrow \infty}K=\frac{8 \left(b^2 \Lambda-1 \right)^2}{3 b^4}.
\end{equation}
\par(iii) For $-1<\beta <0$, the Kretschmann scalar is finite, close to the origin, and diverges for points very far from the black hole, according to
\begin{equation}
\lim_{r\rightarrow 0}K=\frac{8 \left(\Lambda  l^2+3\right)^2}{3 l^4}.
\end{equation}
\begin{equation}
\lim_{r\rightarrow \infty}K=\infty.
\end{equation}
\par(iv) For $0<\beta<2$ and $\beta>2$, the Kretschmann scalar diverges very close to the origin and is finite in a region very far from the black hole.
\begin{equation}
\lim_{r\rightarrow 0}K=\infty.
\end{equation}
\begin{equation}
\lim_{r\rightarrow \infty}K=\frac{8 \Lambda ^2}{3}.
\end{equation}

\subsubsection{For $\beta=2$}

Now, considering the case $\beta=2$, we obtain the following result,
\begin{equation}
\begin{aligned}
K=&R_{\alpha\beta\mu\nu}R^{\alpha\beta\mu\nu}
\\
=&+\frac{6 b^2 \epsilon ^2}{r^6}+\frac{12 b^2 \epsilon ^2 \log ^2(r)}{r^6}+\frac{8 \Lambda ^2}{3}
\\
&-\frac{4 b \epsilon }{l^2 r^3}-\frac{4 b \Lambda  \epsilon }{3 r^3}+\frac{768 l^2 m^3 \left(l^2 m-r^3\right)^3}{\left(2 l^2 m+r^3\right)^6}
\\
&+\frac{96 l^2 m^2 \left(2 \Lambda  l^2 m-3 b \epsilon \right)}{\left(2 l^2 m+r^3\right)^3}
\\
&+\frac{8 m \left(3 b \epsilon +m \left(6-4 \Lambda  l^2\right)\right)}{\left(2 l^2 m+r^3\right)^2}+\frac{4 b \epsilon }{l^2 \left(2 l^2 m+r^3\right)}
\\
&+8 b \epsilon  \log (r) \left(-\frac{b \epsilon }{r^6}-\frac{6 m \left(r^3-4 l^2 m\right)}{\left(2 l^2 m+r^3\right)^3}\right).
\label{eq:1.34}
\end{aligned}
\end{equation}

 \noindent whose limits as $r\rightarrow 0$ and $r\rightarrow \infty$, are the following

\begin{equation}
\lim_{r\rightarrow 0}K=\infty\,\,\text{and}\,\,\lim_{r\rightarrow \infty}K=\frac{8 \Lambda ^2}{3}.
\label{eq:1.36}
\end{equation}

Note that these limits are the same obtained for $0 < \beta < 2$ and $\beta>2$.

These results tell us that the inclusion of the fluid of strings changes the regularity of the Hayward solution for the following values of $\beta$, such as $\beta < -1$ and $\beta > 0$. Otherwise, the regularity of the Hayward black hole solution is preserved in the interval $-1\leq\beta<0$.

\subsection{Geodesics and effective potential}

In order to complete the studies concerning the spacetime's  singularity, let us examine, now, the behavior of geodesics with respect to the completeness or incompleteness.
Firstly, we consider the possible existence of incomplete geodesics, whose proper length is finite and, therefore, a particle moving along such geodesics would experience only a finite interval of existence and, in principle, cannot continue any longer.
In the case of a timelike geodesic, it is natural to consider the proper time experienced by a particle  moving along the geodesic. 

Now consider  Eqs. (\ref{eq:line_element})and (\ref{eq:f_r}), which represent the solution and the metric function,$f(r)$, respectively and assume that $q^3\equiv2l^2m$. 
We will analyze geodesic equations by considering the geodesic action and the variational principle to write a Lagrangian that describes the metric as

\begin{equation}
    2\mathcal{L} = f(r) \dot{t}^2-\frac{1}{f(r)}\dot{r}^2-r^2\dot{\theta}^2-r^2\sin^2\theta\dot{\phi}^2,\label{lagrangian}
\end{equation}
where the “point” means derivative with respect to the proper time, $\tau$. For simplicity, let us restrict the analysis of the geodesics to the equatorial plane of the black hole, $\theta=\frac{\pi}{2}$. Using the Euler-Lagrange equations, we get
\begin{align}
    E &= f(r)\dot{t},
\label{energia}
\end{align}
\begin{align}
    J &= -r^2\dot{\phi},
\label{momento angular}
\end{align}

\noindent where $E$ and $J$ are
constants which correspond to the Killing vectors $\partial_t$ and $\partial_\phi$, respectively. These constants can be interpreted
as the energy $E$ and the angular momentum $J$ of the particle that is moving around the black hole. 

Rescaling the parameter $\tau$, we can define $L=2\mathcal{L}$, which, for time-like geodesics, is equal to $+1$, for space-like geodesics is equal to $-1$ and is equal to $0$ for null geodesics \cite{chandrasekhar1983mathematical}. Substituting Eqs. (\ref{energia}) and (\ref{momento angular}) into Eq. (\ref{lagrangian}), we get

\begin{equation}
\dot{r}^2=E^2-V_{eff},
\label{energia e potencial efetivo}
\end{equation}

\noindent where

\begin{equation}
V_{eff}=f(r)\left(\frac{J^2}{r^2}+L\right).
\label{potenciale efetivo}
\end{equation}

Consider a probe massless particle ($L=0$) that falls radially $(J=0)$ into the black hole. Thus, from Eq. (\ref{potenciale efetivo}), we find that $V_{eff}=0$, and as a consequence, the proper time is finite, which means that the photons will always reach the point $r=0$, following a straight line trajectory, arising from infinity, in a finite time.
According to the result concerning the Kretschmann scalar, this quantity is finite at $r=0$, for $\beta$ in the interval $-1<\beta<0$, and then, the spacetimes are non-singular or regular.
For values of $\beta$ , such the $\beta<-1$ and $\beta>0$, there is a curvature singularity at $r=0$, and as a consequence, the massless particle will disappear at this point. In this case, we conclude that the spacetimes are singular.

Let's now consider a probe massive particle $(L=1)$ that falls radially $(J=0)$ into the black hole. The radial geodesic equation of motion of a test particle is given by
\begin{equation}
    \dot{r}^2 = E^2- f(r),
    \label{eqmovimento}
\end{equation}
while the effective potential reads:
\begin{equation}
    V_{eff} = f(r).
\end{equation}

\begin{figure*}[!ht]
\centering
\begin{subfigure}[!]{0.45\linewidth}
\includegraphics[scale=0.6]{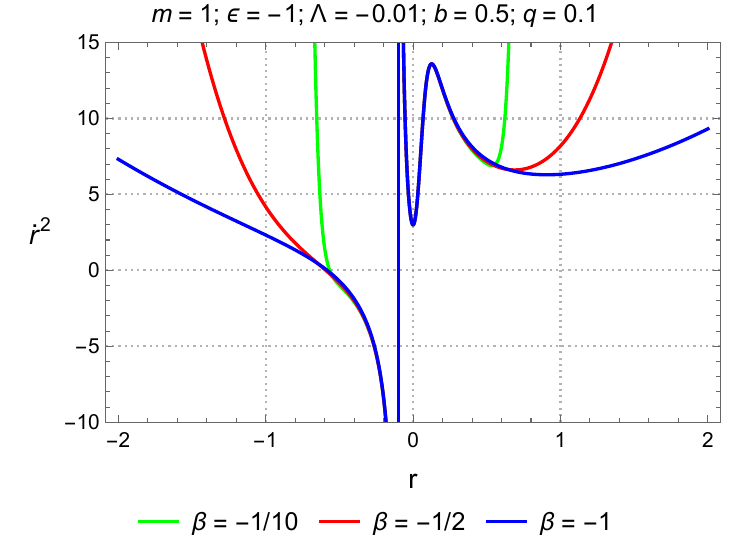}
\caption{}\label{im0a} 
    \vspace{4ex}
      \end{subfigure}
\begin{subfigure}[!]{0.45\linewidth}
\includegraphics[scale=0.6]{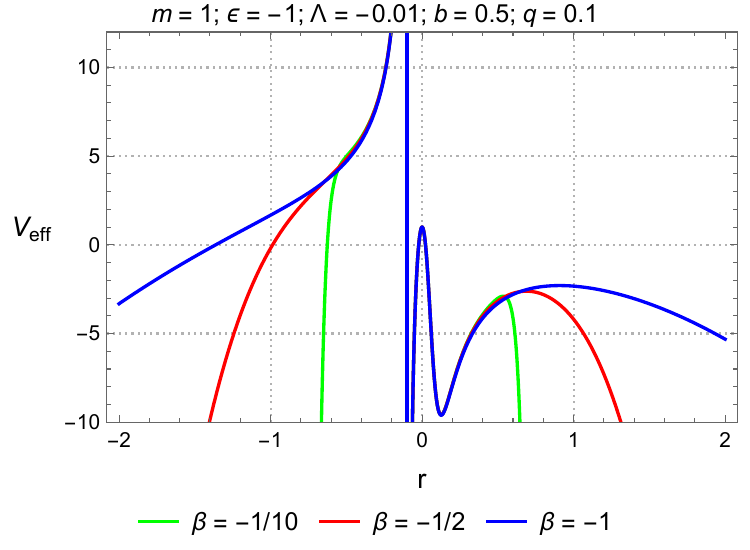}
  \caption{}\label{im0b} 
    \vspace{4ex}
     \end{subfigure}
\begin{subfigure}[!]{0.45\linewidth}
\includegraphics[scale=0.6]{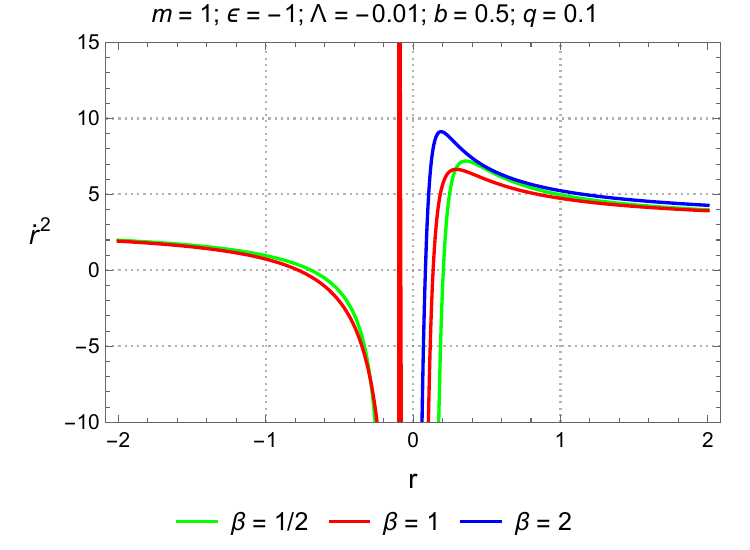}
  \caption{}\label{im0c} 
        \end{subfigure}
\begin{subfigure}[!]{0.45\linewidth}
\includegraphics[scale=0.6]{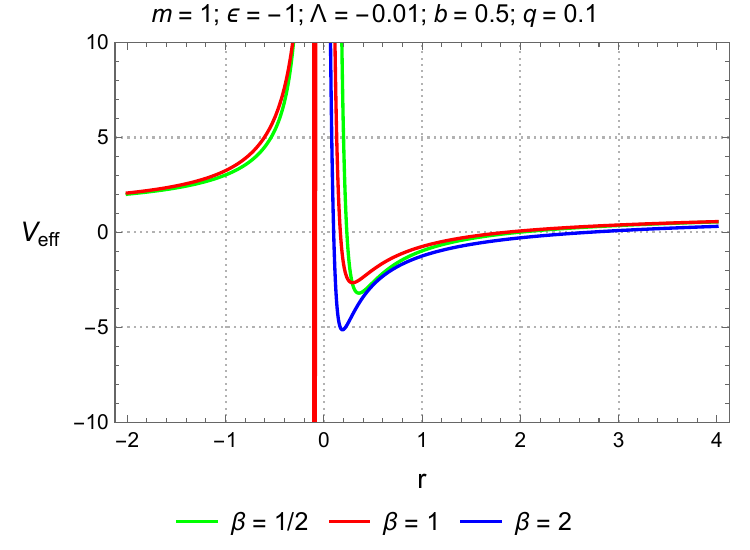}
  \caption{}\label{im0d} 
        \end{subfigure}
  \caption{The left plot shows $\dot{r}^2$ for $E=2$. The right plot shows the effective potential $V_{eff} = f(r)$.} 
\label{im0}
\end{figure*}

Using the relation

\begin{equation}
\begin{aligned}
  &\frac{dr}{dt}\frac{dt}{d\tau}=\frac{dr}{d\tau}\\
  &\left(\frac{dr}{dt}\right)^2\left(\frac{dt}{d\tau}\right)^2=\left(\frac{dr}{d\tau}\right)^2\Rightarrow\left(\frac{dr}{dt}\right)^2\dot{t}^2=\dot{r}^2,
\end{aligned}
\label{eq:1.85.1}
\end{equation}

\noindent into Eq.(\ref{eqmovimento}) and using Eqs.(\ref{energia}), we get

\begin{equation}
\left(\frac{dr}{dt}\right)^2=f(r)^2\left(1-\frac{f(r)}{E^2}\right).
\label{eq:1.86.1}
\end{equation}

Substituting Eq.(\ref{eq:f_r}) into Eq.(\ref{eq:1.86.1}), we can find the relationship between the coordinates $t$ and  $r$ for the radial movement of the particle:

\begin{equation}
\pm t=\int\frac{dr}{\sqrt{f(r)^2-\frac{f(r)^3}{E^2}}}.
\label{eq:1.91.1}
\end{equation}

From Eq.(\ref{eqmovimento}), we get the relationship between the proper time $\tau$ and the radial coordinate $r$:

\begin{equation*}
\left(\frac{dr}{d\tau}\right)^2=E^2-f(r),
\end{equation*}

\begin{equation}
\pm\tau=\int\frac{dr}{\sqrt{E^2-f(r)}}.
\label{eq:1.92.1}
\end{equation}

In order to analyze the regularity or not of the solution, we need to obtain the value of the integral given by Eq. (\ref{eq:1.92.1}) for a geodesic reaching the point $r=0$. So, if we consider $r \approx 0$, we get

\begin{equation}
\pm\tau \approx \int \left[ (\beta\epsilon (2-\beta)^{-1})^{1/2}\left(\frac{b}{r}\right)^{1/\beta}\right]^{-1}dr,
\label{eq:1.92.3}
\end{equation}
whose approximate solution is 
\begin{equation}
\pm\tau \approx \frac{\beta b^{-1/\beta } r^{1+\frac{1}{\beta }}}{(\beta +1) \sqrt{\frac{\beta \epsilon }{2-\beta}}}
\label{eq:1.92.3.1}
\end{equation}

This result tells us that the solution is proportional to $r^{1+1/\beta}$, which means that the integral does not diverge for $-1<\beta<0$. So, we can conclude that the spacetime is regular for $-1<\beta<0$. Proceeding analogously, it is possible to verify that the spacetime is singular for $\beta = 2$ and for values of $\beta$ such that
$\beta < -1$ and $\beta >0$.

Now, let us consider Eq. (\ref{eq:1.92.1}), for time-like radial geodesics described by a massive particle that falls radially, in which case we have that $f(r)=V_{eff}$. Thus, if we consider a particle with an energy below the maximum of the effective potential barrier, then it will hit the barrier at some value of the radial coordinate and roll back.
On the other hand, if this energy is higher than this maximum, it will reach the center in a finite time.
In this case, the particle will hit the singularity for any value of $\beta>0$ and $\beta<-1$, and as a consequence, these spacetimes are geodesically incomplete. With respect to the spacetimes corresponding to $-1<\beta<0$, they are geodesically complete. 

Let us examine these questions using the panels in Fig.(\ref{im0}). Note that for $-1\leq\beta<0$, Fig. (\ref{im0b}), particles with sufficient energy reach the top of the barrier and still have enough kinetic energy, travel in all spacetime, which means that the spacetime is geodesically complete. 
In the case $\beta>0$, Fig. (\ref{im0d}), the effective potential tends to an infinitely large value and the particle will roll back, which means that the spacetime will not be covered and, therefore, the geodesics are incomplete.
 
In Fig.(\ref{im0b}), the probe particle can overcome the potential barrier and reach the point $r=0$ in a finite time. This fact, from the point of view of the characteristics of the spacetime with respect to singularity, is exactly what is predicted by the values of the Kretschmann scalar, which are finite for these values of $\beta$. 

In Fig.(\ref{im0d}), we can conclude that the probe particle cannot reach the point $r=0$ in a finite time. This indicates that the geodesics are incomplete and, therefore, the spacetime is singular. This conclusion is confirmed by the fact that for $\beta>0$, the Kretschmann scalar is infinite when $r\rightarrow 0$.

Among the cases considered, only the cases in which $\beta$ is in the interval $-1\leq\beta<0$ there is no curvature divergence at $r=0$, which indicates that the spacetime is regular, for different values of $\beta$, in this interval, taking into account the criterion based on the curvature scalar invariants to examine the existence of singularities of the metric.
On the other hand, when $\beta$ assumes values outside this interval, the spacetime is singular.

In summary, the analysis of the geodesics with the aim of knowing the characteristics of the spacetimes in relation to the singularity, confirms what is predicted by the results provided by the Kretschmann scalar, at least in the cases considered.

\section{Thermodynamics}
\label{sec:therm}

The first law of black hole thermodynamics for charged and static black holes is given by \cite{gunasekaran2012extended}

\begin{equation}
    dM = T_HdS + \Phi_H dq +V_H dP, \label{eq:first_law}
\end{equation}

\noindent where $M$ is the mass of black hole, $\Phi_H$ is the magnetic potential, $P$ is the thermodynamic pressure, written as \cite{kubizvnak2012p}

\begin{equation}
    P = -\frac{\Lambda}{8\pi}. \label{pressure}
\end{equation}

\noindent $S$ is the entropy of the system, which is related to the area of the black hole horizon $A$ by the equation \cite{bekenstein1973black}

\begin{equation}
    S = \frac{A}{4} = \pi r^2_+.\label{eq:entropy}
\end{equation}

From the first law, we can obtain the temperature, the magnetic potential, and the volume as follows:

\begin{equation}
    T_H = \frac{\partial M}{\partial S},\,\,\,\,\Phi_H = \frac{\partial M}{\partial q}\label{eq:TH_Ph}\,\,\text{and}\,\,V_H = \frac{\partial M}{\partial P}.
\end{equation}

It is worth calling attention to two considerations that we will assume in this work: 

(i) We will focus on the solution with $-1<\beta<0$, since in this interval the metric obtained remains regular at the origin, maintaining the same characteristic as Hayward's solution.; 
\par
(ii) Let's also consider the following redefinition $q^3\equiv2l^2M$. Once $q^3$ is fixed, we are automatically fixing the product $ml^2$ and not the mass itself. This can be done without any problem, it is a perfectly valid identification. If we don't redefine the parameter, we'll have an ambiguity in the sense that we'll have two possible pressure values, one of which corresponds to a situation that physically doesn't reflect reality.

So, with these considerations, the Eq. (\ref{eq:f_r}) can be written as

\begin{align}
f(r)&=1-\frac{2 M r^2}{r^3+q^3}-\frac{\Lambda  r^2}{3}+\epsilon\beta(\beta -2)^{-1} \left(\frac{b}{r}\right)^{2/\beta}.
\label{f_r}
\end{align}

We can obtain $M$ from the condition $f(r_+) = 0$. 

\begin{equation}
\begin{aligned}   
M=&\frac{3 \beta\epsilon  \left(\frac{b}{r_+}\right){}^{2/\beta }-\Lambda\beta r_+^2+3\beta+2 \Lambda  r_+^2-6}{6 (\beta -2) r_+^2\left(q^3+r_+^3\right)^{-1}}.
\label{eq:mass}
\end{aligned}
\end{equation}

\begin{figure*}[!ht]
\centering
\begin{subfigure}[!]{0.45\linewidth}
\includegraphics[scale=0.55]{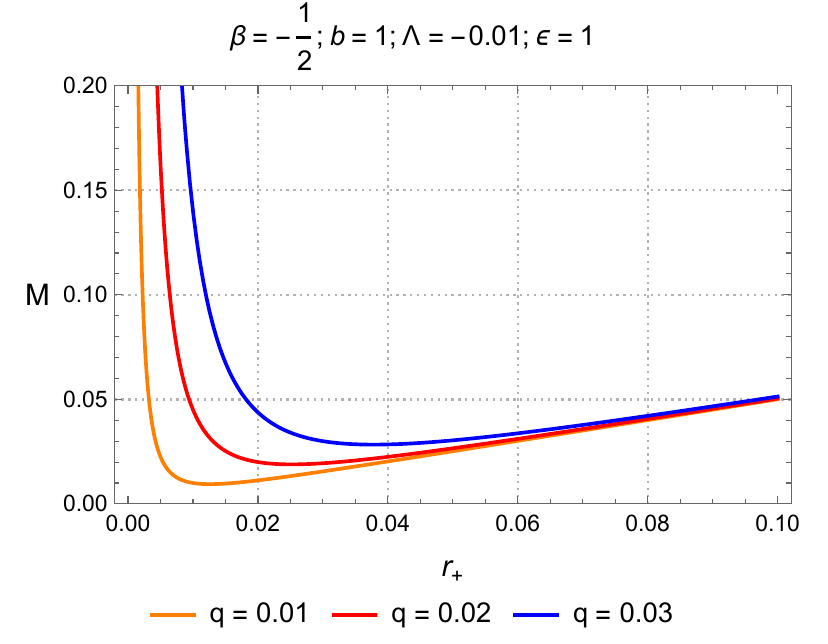}
\caption{}\label{ma} 
    \vspace{4ex}
      \end{subfigure}
\begin{subfigure}[!]{0.45\linewidth}
\includegraphics[scale=0.55]{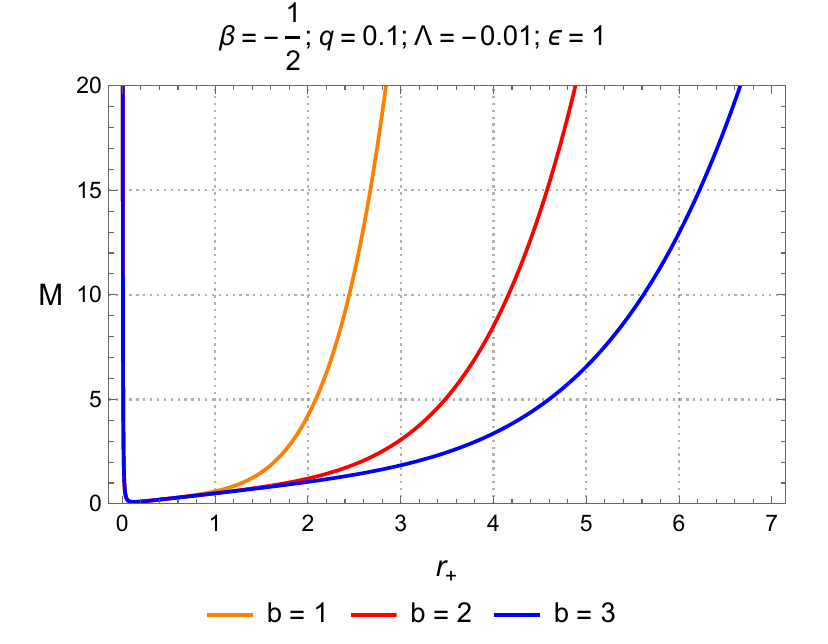}
  \caption{}\label{mb} 
    \vspace{4ex}
     \end{subfigure}
\begin{subfigure}[!]{0.45\linewidth}
\includegraphics[scale=0.55]{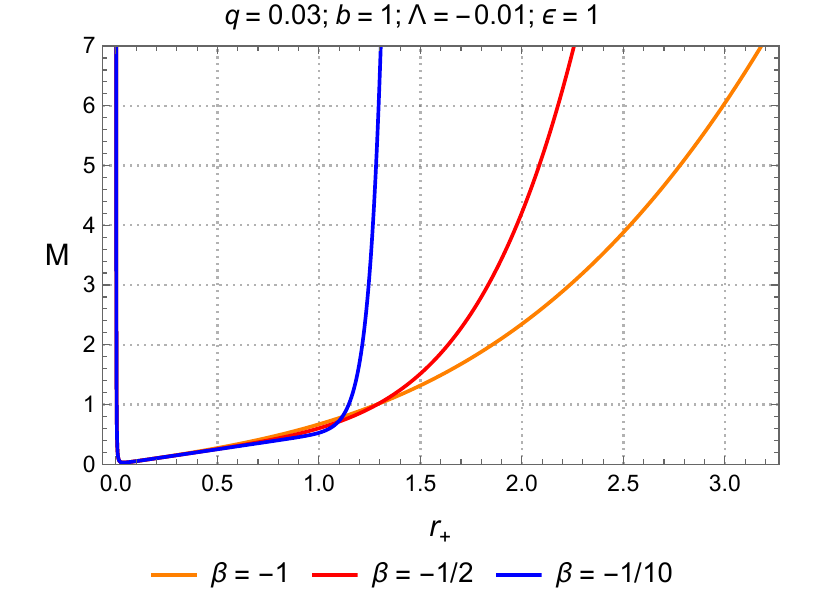}
  \caption{}\label{mc} 
        \end{subfigure}
        \begin{subfigure}[!]{0.45\linewidth}
\includegraphics[scale=0.55]{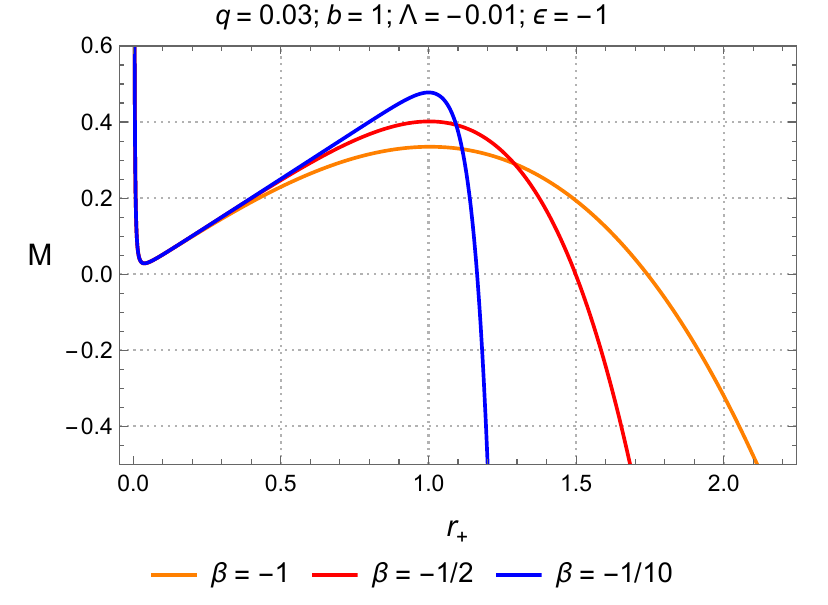}
  \caption{}\label{md} 
        \end{subfigure}
  \caption{Black hole mass $M$ as a function of the radius of the event horizon $r_{+}$ for different values of $q$, $b$, $\epsilon$ and $\beta$.} 
\label{mab}
\end{figure*}

In Fig. (\ref{ma}), we can observe the behavior of the mass $M$ for different values of $q$ which indicates the intensity of the Hayward parameter. We can notice that, if we fix a value to $r_{+}=0.02$, the mass increases as the $q$ parameter increases. In Fig. (\ref{mb}) the behavior of the mass is shown in terms of the parameter $b$, which indicates the intensity of the fluid of strings. Here,  we can notice that, if we fix a value to $r_{+}=2$, the mass increases as the $b$ parameter decreases. In the two previous situations, the mass has positive values for $r_{+}>0$.

In Figs. (\ref{mc})-(\ref{md}), we can observe the behavior of the mass $M$, for different values of $\beta$ and $\epsilon$, as a function of the radius of the event horizon $r_+$. Here, we can notice that for different values of $\beta$, the mass of black holes has the same limits.  It is worth calling attention to the fact that in the analysis of the mass, we fixed the values of the parameter $\beta$, in such a way that the solutions are regular.

Now, we can calculate the temperature $T_H$ using Eq. (\ref{eq:entropy}), Eq. (\ref{eq:TH_Ph}) and Eq. (\ref{eq:mass}), which results in

\begin{equation}
  \begin{aligned}
  T_H=&\frac{\left(-\frac{(\beta +1) \epsilon  b^{2/\beta } r_+^{-2/\beta }}{\beta -2}-1\right)}{2 \pi q^{-3}  r_+^4}+\frac{\epsilon  b^{2/\beta } r_+^{3-\frac{2}{\beta }}-\Lambda  r_+^5+r_+^3}{4 \pi  r_+^4}.
\end{aligned}
\end{equation}

Hawking showed that black holes emit
radiation and its corresponding temperature (known as Hawking temperature) is given by  \cite{hawking1975particle}

\begin{equation}
    T_{\kappa}  = \frac{\kappa}{2\pi},
    \label{eq:Tk}
\end{equation}

\noindent for stationary space-times. The surface gravity is defined by \cite{wald2001thermodynamics}

\begin{equation}
    \kappa = \frac{f'(r)}{2}\Bigg{|}_{r=r_+},
\end{equation}

\noindent with $'$ denoting a derivative concerning the radial coordinate and the $r_+$ being the event horizon radius. Now, considering Eq.(\ref{f_r}) and using the mass of the black hole given by Eq.(\ref{eq:mass}), we obtain:

\begin{equation}
\begin{aligned}
    T_{\kappa} =&-\frac{\epsilon  \left(\frac{b}{r_+}\right){}^{2/\beta }}{2 \pi  (\beta -2) r_+}-\frac{\Lambda  r_+}{6 \pi }
    \\
    &+\frac{3 \beta \epsilon  \left(\frac{b}{r_+}\right){}^{2/\beta }+3 \beta -\text{$\beta $r}_+^2 \Lambda +2 \Lambda  r_+^2-6}{12 \pi  (\beta -2) r_+ \left(q^3+r_+^3\right)(r_+^3-2 q^3)^{-1}}.
\end{aligned}
\label{T_k}
\end{equation}

We can notice that the temperature $T_\kappa$ is different from $T_H$. In addition to the temperature, the other quantities derived from the first law of thermodynamics also present problems. To solve this problem, the first law of black hole thermodynamics needs to be modified \cite{ma2014corrected,maluf2018thermodynamics,singh2020thermodynamics,rodrigues2022bardeen}.

The usual first law arises in a context where the Lagrangian of the theory does not explicitly depend on the mass of the black hole. Thus, when building the first law for solutions with nonlinear electrodynamics, derivatives of the stress-energy tensor must be taken into account and these corrections must modify the first law in such a way that the old thermodynamic quantities must relate to the new ones through a correction factor. The new first law is written as \cite{ma2014corrected,maluf2018thermodynamics,singh2020thermodynamics,rodrigues2022bardeen}.

\begin{equation}
    d\mathcal{M}= TdS + \Phi dq + VdP,
\end{equation}

\noindent where

\begin{equation}
    d\mathcal{M}=WdM,
\end{equation}

\noindent and $W$ is the correction factor, given by

\begin{equation}
W=\left(1+4\pi\int^{\infty}_{r_+}r^2\frac{\partial T^0_0}{\partial M}dr \right),
\end{equation}

\noindent and $T^0_0$ is one of the components of the stress-energy tensor.

Now for the case of the Hayward black hole with cosmological constant and a fluid of strings, the relation between the temperatures, magnetic potential, and volume are

\begin{equation}
\begin{aligned}
    T_\kappa &= WT_H = W \frac{\partial M}{\partial S},\\
    \Phi &= W \Phi_H = W \frac{\partial M}{\partial q},\\
    V &= W V_H = W \frac{\partial M}{\partial P},
\label{TV}
\end{aligned}
\end{equation}

\noindent where the factor $W$ is given by

\begin{equation}
    W = W(r_+,q) = \frac{r_+^3}{q^3+r_+^3}.
    \label{correction factor}
\end{equation}

\begin{figure*}[!ht]
\centering
\begin{subfigure}[!]{0.48\linewidth}
\includegraphics[scale=0.65]{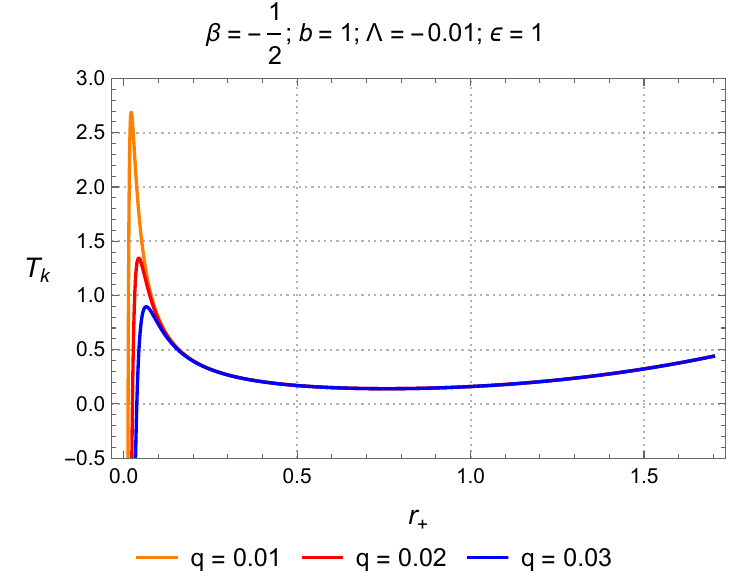}
\caption{}\label{Ta} 
    \vspace{4ex}
      \end{subfigure}
\begin{subfigure}[!]{0.48\linewidth}
\includegraphics[scale=0.65]{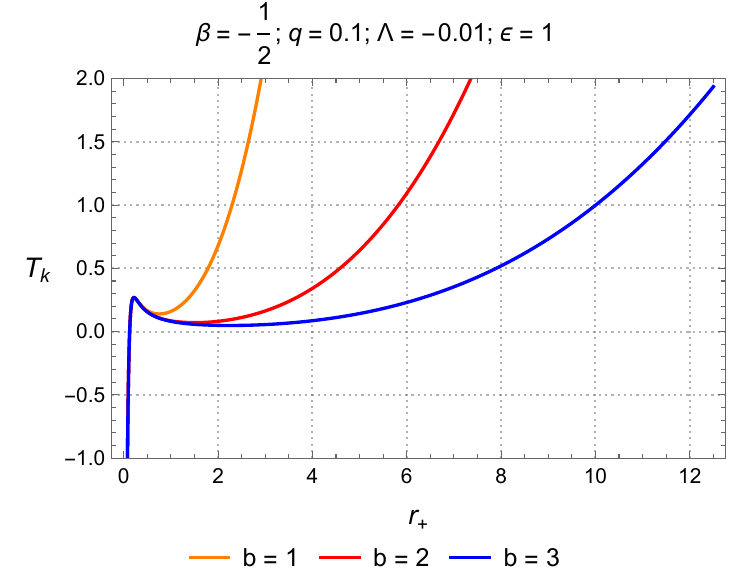}
  \caption{}\label{Tb} 
    \vspace{4ex}
     \end{subfigure}
\begin{subfigure}[!]{0.48\linewidth}
\includegraphics[scale=0.65]{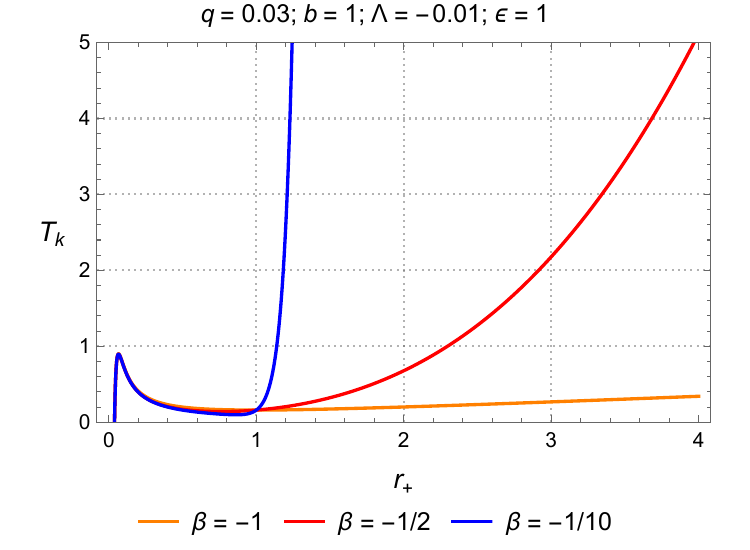}
  \caption{}\label{Tc} 
        \end{subfigure}
        \begin{subfigure}[!]{0.48\linewidth}
\includegraphics[scale=0.65]{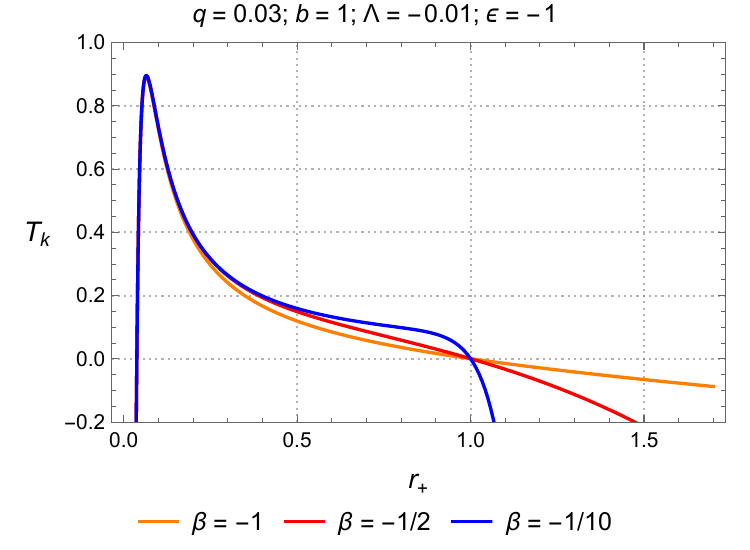}
  \caption{}\label{Td} 
        \end{subfigure}
  \caption{Black hole temperature $T_k$ as a function of the radius of the event horizon $r_{+}$ for different values of $q$, $b$, $\epsilon$ and $\beta$.} 
\label{Tab}
\end{figure*}

In Fig. (\ref{Ta}), we can observe the behavior of the temperature $T_k$ for different values of $q$ which indicates the intensity of the Hayward parameter. We can notice that, if we fix a value to $r_{+}=0.1$, the temperature increases as the $q$ parameter decreases. In Fig. (\ref{Tb}) the behavior of the temperature is shown in terms of the parameter $b$, which indicates the intensity of the fluid of strings. Here,  we can notice that, if we fix a value to $r_{+}=2$, the temperature increases as the $b$ parameter decreases. Also note that in Fig. (\ref{Tb}) the maximum value of the temperature doesn't change for different values of $b$, but the maximum of $r_+$ depends on the values of $b$.

In Figs. (\ref{Tc})-(\ref{Td}), we can observe the behavior of the temperature $T_k$, for different values of $\beta$ and $\epsilon$, as a function of the radius of the event horizon $r_+$. Here, we can notice that for different values of $\beta$, the temperature of black holes has the same limits. It is worth calling attention to the fact that in the analysis of the temperatures, we fixed the values of the parameter $\beta$, in such a way that the solutions are regular.

\subsection{Heat capacity}

Heat capacity provides information about the thermodynamic stability of a system. 
It is important to distinguish the form as the heat is received in the system. If the volume is constant, we have an isochoric thermal capacity $C_v$. If the pressure is constant, we have an isobaric thermal capacity $C_P$.

From classic thermodynamics, we know that the stability of a thermodynamic system requires $C_P\geq C_v\geq0$. 
 Now from the thermodynamics of black holes, and considering the case of pressure constant, we have \cite{davies1977thermodynamic}
\begin{equation}
    C_P = T_\kappa\frac{\partial S}{\partial T_\kappa}\bigg|_P.
\end{equation}
We can also write this heat capacity in terms of the correction factor $W(r_+,q)$, given by
\begin{equation}
    C_P = W(r_+,q)\frac{\partial T_H}{\partial T_\kappa}\Tilde{C}_P,
\end{equation}
where $\Tilde{C}_P = T_H\frac{\partial S}{\partial T_H}$ is the heat capacity defined in terms of the usual first law. So the heat
capacity of the Hayward black hole with cosmological constant and surrounded by a fluid of strings is
\begin{equation}
\begin{aligned}
C_P=-\frac{2 \beta  S \left(\psi _1+\psi _2\right) \left(\pi ^{3/2} q^3+S^{3/2}\right)}{\pi ^{\frac{1}{\beta }+1} \left(\psi _3+\psi _4\right) \epsilon  \left(\frac{b}{\sqrt{S}}\right)^{2/\beta }+(\beta -2) \beta  \left(\psi _5+\psi _6\right)},
\end{aligned}
\label{heat capacity}
\end{equation}

\noindent where

\begin{equation*}
  \psi _1=\pi ^{\frac{1}{\beta}+1} \epsilon  \left(\frac{b}{\sqrt{S}}\right)^{2/\beta} \left(2 \pi ^{3/2} (\beta +1) q^3-(\beta -2) S^{3/2}\right),
\end{equation*}
\begin{equation*}
    \psi _2=(\beta -2) \left(2 \pi ^{5/2} q^3+\Lambda  S^{5/2}-\pi  S^{3/2}\right),
\end{equation*}
\begin{equation*}
    \psi _3=2 \pi ^3 (\beta +1) (\beta +2) q^6-\left(\beta ^2-4\right) S^3,
\end{equation*}
\begin{equation*}
    \psi _4=2 \pi ^{3/2} (\beta  (5 \beta +3)+4) q^3 S^{3/2},
\end{equation*}
\begin{equation*}
    \psi _5=2 \pi ^4 q^6+10 \pi ^{5/2} q^3 S^{3/2}-\pi  S^3,
\end{equation*}
\begin{equation*}
    \psi _6=-4 \pi ^{3/2} \Lambda  q^3 S^{5/2}-\Lambda  S^4,
\end{equation*}

\begin{figure*}[!ht]
\centering
\begin{subfigure}[!]{0.48\linewidth}
\includegraphics[scale=0.65]{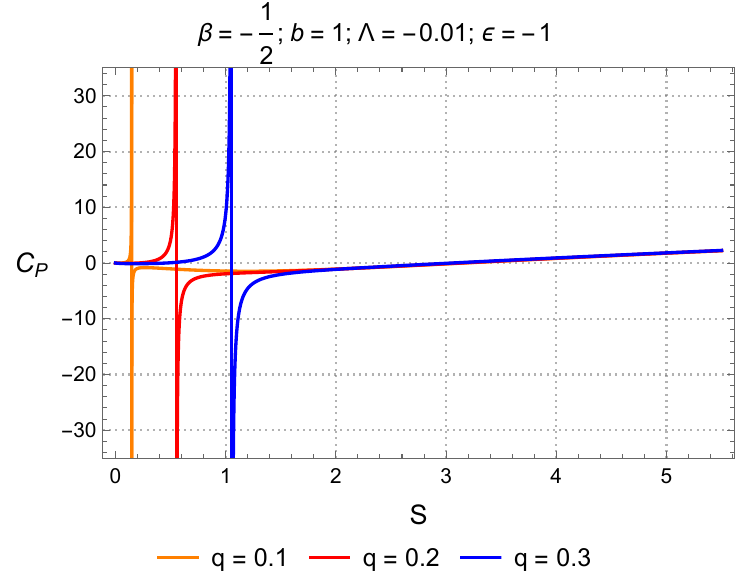}
\caption{}\label{Ca} 
    \vspace{4ex}
      \end{subfigure}
\begin{subfigure}[!]{0.48\linewidth}
\includegraphics[scale=0.65]{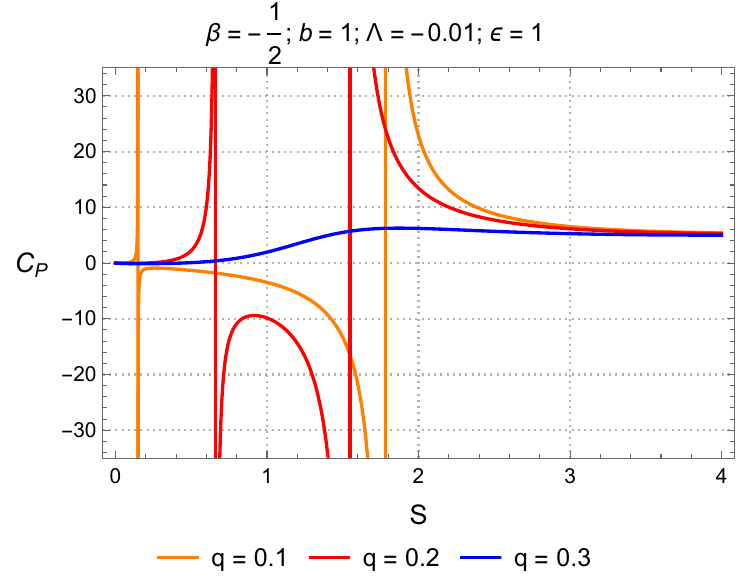}
  \caption{}\label{Cb} 
    \vspace{4ex}
     \end{subfigure}
\begin{subfigure}[!]{0.48\linewidth}
\includegraphics[scale=0.65]{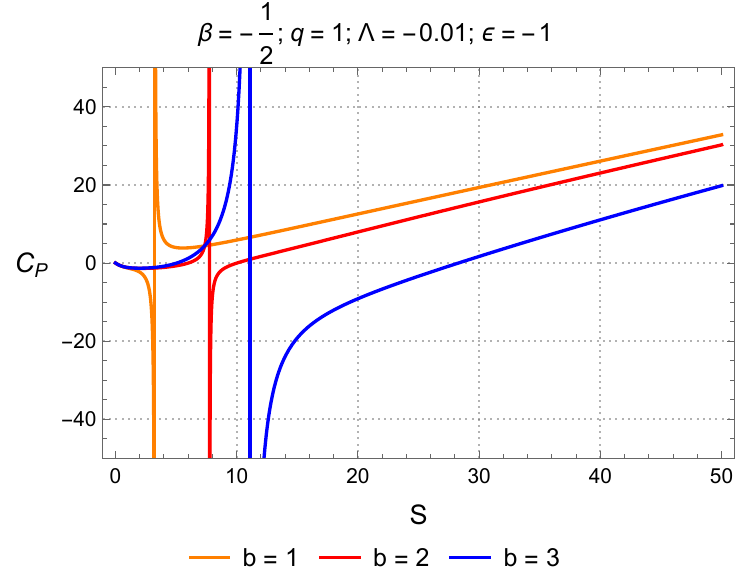}
  \caption{}\label{Cc}
   \vspace{4ex}
        \end{subfigure}
        \begin{subfigure}[!]{0.48\linewidth}
\includegraphics[scale=0.65]{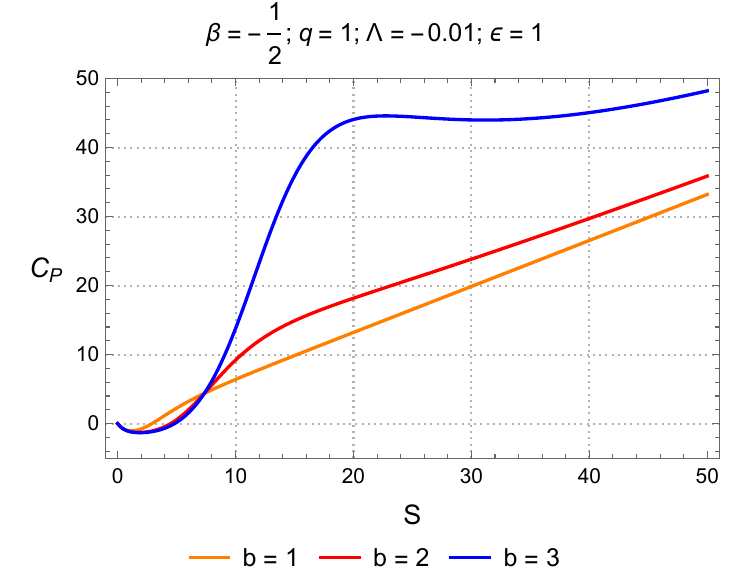}
  \caption{}\label{Cd} 
   \vspace{4ex}
        \end{subfigure}
        \begin{subfigure}[!]{0.48\linewidth}
\includegraphics[scale=0.65]{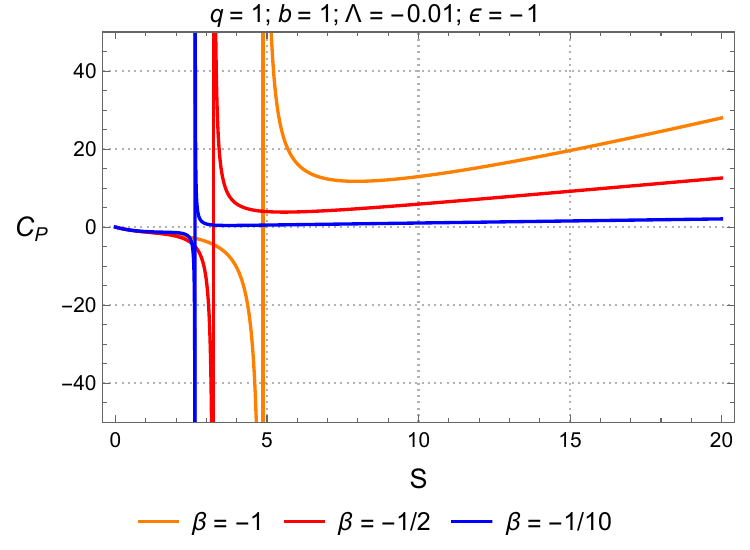}
  \caption{}\label{Ce} 
        \end{subfigure}
        \begin{subfigure}[!]{0.48\linewidth}
\includegraphics[scale=0.65]{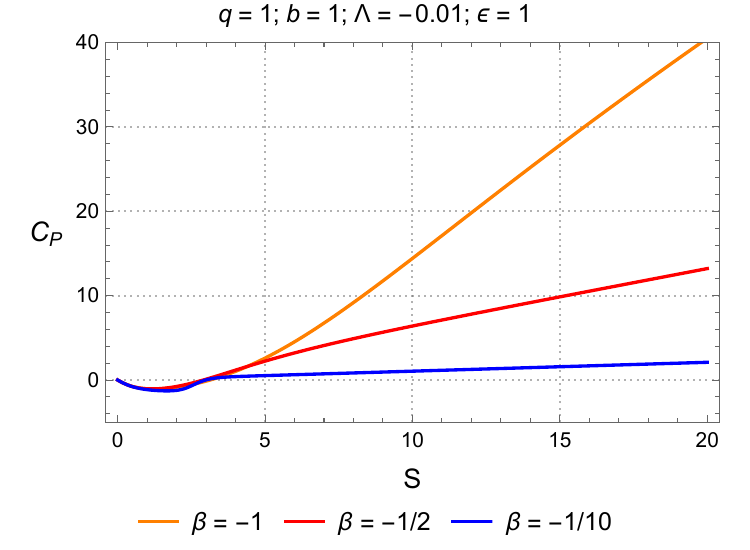}
  \caption{}\label{Cf} 
        \end{subfigure}
  \caption{Heat capacity at constant pressure $C_P$ for different values of $q$, $b$, $\epsilon$ and $\beta$.} 
\label{Capacity}
\end{figure*}

\noindent whose behavior is shown in Fig. (\ref{Capacity}), as a function of the black hole entropy, for different values of the parameters of the Hayward and fluid of strings.

We can conclude that there are values of $S$ for which the heat capacity is positive, as well as there are values for which the heat capacity is negative. In other words, the black hole can be thermodynamically stable or unstable and this stability is related to the
values of the $q$ and $b$.
We can see that the transition point, in which the heat capacity diverges, changes when we
vary these parameters.
Another result is demonstrated in Fig.(\ref{Cb}), that for $q=0.1$ or $q=0.2$, the heat capacity shows two discontinuities, and for $q=0.3$ has only one discontinuity. 

\subsection{Equation of state and critical thermodynamics variables}

In the extended phase space, one can treat the cosmological
constant as thermodynamic pressure and its conjugate quantity as
thermodynamic volume. We will therefore use the following equations (\ref{pressure}), (\ref{TV})-(\ref{correction factor}).

With Eqs. (\ref{pressure}) and (\ref{eq:mass}), the mass can be re-expressed as
\begin{equation}
\begin{aligned}
M=\frac{3 \beta  \epsilon  \left(\frac{b}{r}\right)^{2/\beta }+(\beta -2) \left(8 \pi  P r^2+3\right)}{6 (\beta -2) r^2(q^3+r^3)^{-1}}.
\end{aligned}
\label{new mass}
\end{equation}
Utilizing Eqs. (\ref{TV})-(\ref{correction factor}) and (\ref{new mass}), one can obtain the thermodynamic volume as
\begin{equation}
\begin{aligned}
    V=W(r_+,q)\frac{\partial M}{\partial P}=\frac{4 \pi  r_+^3}{3}.
\end{aligned}
\label{Volume}
\end{equation}

Now let us define $v=2r_{+}$ \cite{li2014effects}, which is double the horizon radius and differs from the usual definition of specific volume in thermodynamics. Thus, substituting $r_{+}=v/2$ and $\Lambda=-8\pi P$ in Eq. (\ref{T_k}) and isolating $P$ in the resultant equation, we get the equation of state::

\begin{equation}
\begin{aligned}
P(T,v)=&\frac{T \left(8 q^3+v^3\right)}{v^4}-\frac{v^3-16 q^3}{2 \pi  v^5}-
\\
&\frac{2^{\frac{2}{\beta }-1} \epsilon  \left(\frac{b}{v}\right)^{2/\beta } \left[(\beta -2) v^3-16 (\beta +1) q^3\right]}{\pi  (\beta -2) v^5}.
\end{aligned}
\label{PTV}
\end{equation}

\begin{figure}[h!]
     \centering
    \includegraphics[width=0.48\textwidth]{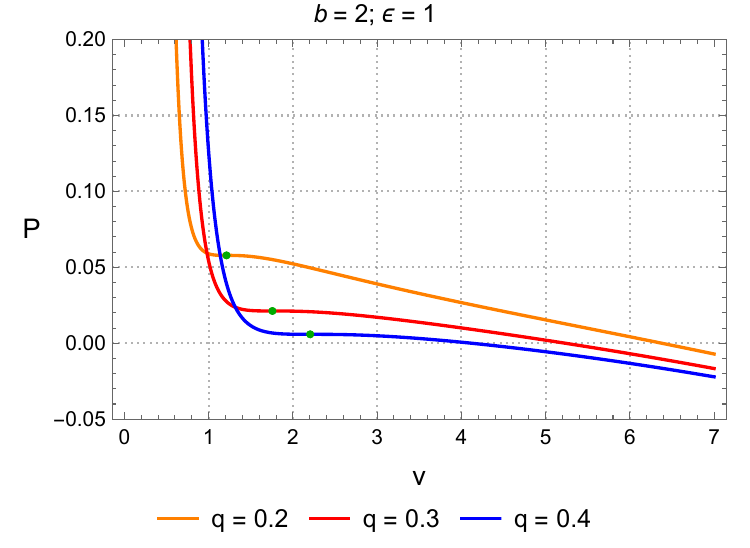}
    \caption{Graphic representation of the $P - v$ diagram
phase for different values of $q$.}
    \label{pv2}
\end{figure}

\begin{figure}[h!]
     \centering
    \includegraphics[width=0.48\textwidth]{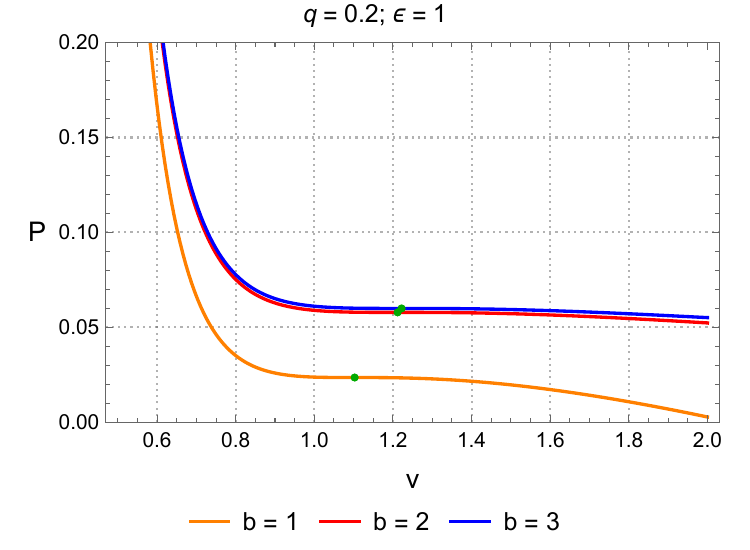}
    \caption{Graphic representation of the $P - v$ diagram
phase for different values of $b$.}
    \label{pv3}
\end{figure}

\begin{figure}[h!]
     \centering
    \includegraphics[width=0.48\textwidth]{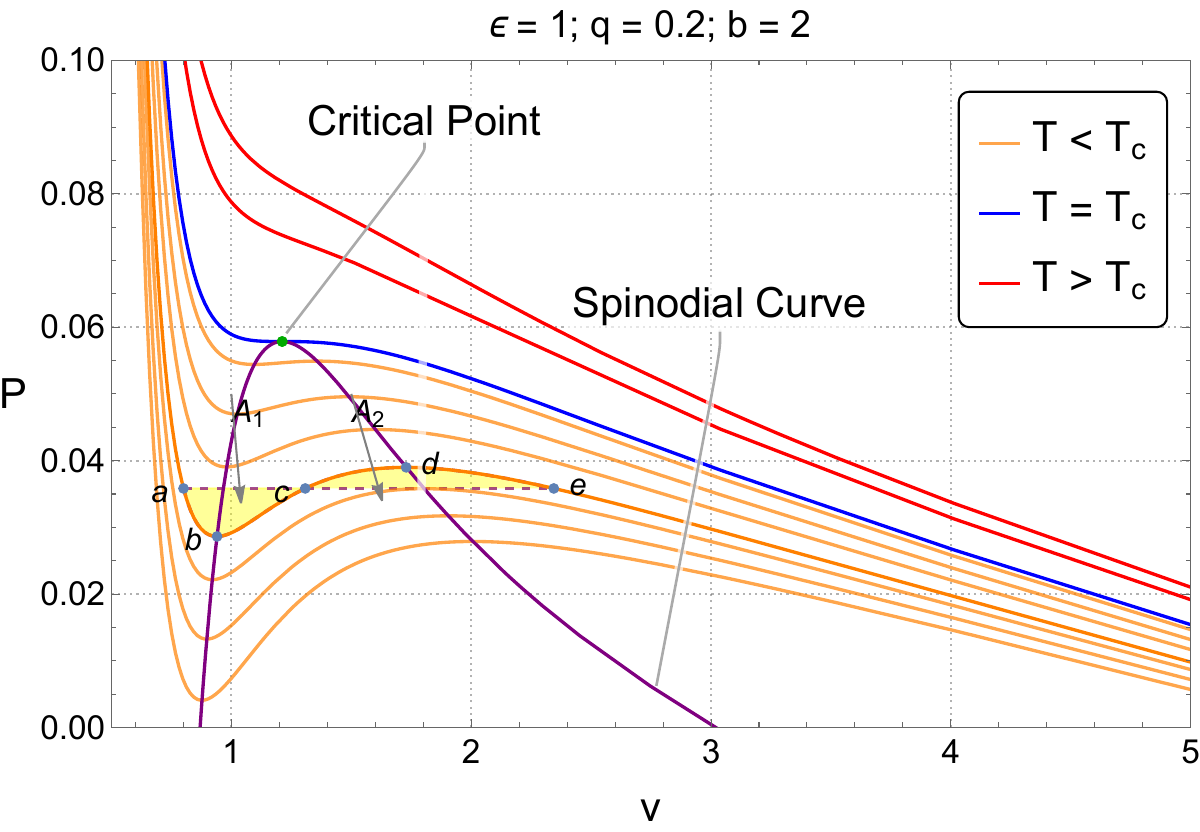}
    \caption{Graphical representation of the equation of state for $b = 2$ and $q = 0.2$. The blue line is the critical isotherm; the orange lines are the isotherms with temperatures below the critical temperature; and the red lines are the isotherms with temperatures above the critical temperature. Additionally, the purple solid line is the spinodal curve, and the purple dashed line is the line of coexistence between phases.}
    \label{pv4}
\end{figure}

The corresponding “$P-v$ diagram” is depicted in Figs. (\ref{pv2})-(\ref{pv4}). In Fig. (\ref{pv4}), we show the isotherm considering three cases, $T < T_c$, $T=T_c$ and $T > T_c$, where $T_c$ is the critical temperature. It is clear that the pressure is higher for higher temperature values. In Figs. (\ref{pv2})-(\ref{pv3}), the green dot indicates the critical point of each curve, whose values are described in the Table (\ref{table:critical_val_q}) and (\ref{table:critical_val_b}). In Fig.(\ref{pv2}), note that the critical values of the pressure decrease when the values of $q$ increase. In Fig.(\ref{pv3}), we see that when the values of $b$ increase, the values of the critical pressure increase slightly.

To obtain the critical points, let's consider the following conditions:
\begin{equation}
    \frac{\partial P}{\partial v}\bigg|_{T=T_c} = 0\,\,\,\text{and}\,\,\,\frac{\partial^2 P}{\partial v^2}\bigg|_{T=T_c} = 0.\label{eq:criticality}
\end{equation}

 Using the first condition, we get the following relation between the critical Hawking temperature, $T_c$, and the critical  specific volume, $v_c$,
\begin{equation}
\begin{aligned}
    T_c=&\frac{v_c^3-40 q^3}{\pi  v_c \left(v_c^3+32 q^3\right)}+
    \\
    &\frac{4^{1/\beta } \epsilon  \left(\frac{b}{v_c}\right){}^{2/\beta } \left[(\beta -2) v_c^3-8 (5 \beta +2) q^3\right]}{\pi  (\beta -2) \beta(\beta +1)^{-1}  v_c \left(v_c^3+32 q^3\right)}.
\end{aligned}
\label{T_c}
\end{equation}

Using Eqs. (\ref{eq:criticality}) and (\ref{PTV}), we arrive at the results

\begin{equation}
\begin{aligned}
    &+\frac{(\beta -2) \beta ^2 \left(224 q^3 v_c^3-v_c^6+1280 q^6\right)}{\pi  (\beta -2) \beta ^2 v_c^7 \left(v_c^3+32 q^3\right)}
    \\
    &+\frac{4^{1/\beta } (\beta +1) \epsilon  \left(\frac{b}{v_c}\right){}^{2/\beta }}{\pi  (\beta -2) \beta ^2 v_c^7 \left(v_c^3+32 q^3\right)}\left[256 (\beta +2) (5 \beta +2) q^6\right.
    \\
    &\left.\left.+16 (\beta  (14 \beta -3)+10) q^3 v_c^3-\left(\beta ^2-4\right) v_c^6\right)\right]=0,
\end{aligned}
\label{v_c}
\end{equation}

\noindent for the critical parameters for temperature, specific volume, and pressure, $T_c, v_c$, and $p_c$, respectively. However, these values are only found for numeric solutions. In the Tables (\ref{table:critical_val_q}) and (\ref{table:critical_val_b}) the values for the critical temperature, specific volume, and pressure
are shown in two different situations, one of which considers different values of $q$ and the other for different values of $b$, in both cases with $\beta = -1/2$.

\begin{table}[ht]
\caption{Behaviour of the critical parameters $v_c$, $T_c$, $P_c$ and $\frac{P_c v_c}{T_c}$ for different values of $q$ and for $\epsilon = 1$ and $b = 2$}.

\begin{tabular}{clllclllclllclllc}
\hline
$q$  &  $v_c$    &  $T_c$     &  $p_c$      &  $\frac{p_c v_c}{T_c}$ \\\hline
0.2 &  1.211940 &  0.186423  &  0.057860   &  0.376153            \\
0.3  &  1.756850 &  0.119386  &  0.021255   &  0.312787             \\
0.4    &  2.205190  &  0.081176 &  0.005885 &  0.159868              \\\hline
\end{tabular}
\label{table:critical_val_q}
\end{table}
\begin{table}[ht]
\caption{Behaviour of the critical parameters $v_c$, $T_c$, $P_c$ and $\frac{P_c v_c}{T_c}$ for different values of $b$ and for $\epsilon = 1$ and $q = 0.2$}
\begin{tabular}{clllclllclllclllc}
\hline
$b$  &  $v_c$    &  $T_c$     &  $p_c$      &  $\frac{p_c v_c}{T_c}$ \\ \hline
1 &  1.102600  &  0.162354  &  0.023540   &  0.159868              \\
2  &  1.211940 & 0.186423  &  0.057860   &  0.376153              \\
3    &  1.222010  &  0.187986 &  0.059939 &  0.389641              \\\hline
\end{tabular}
\label{table:critical_val_b}
\end{table}

With the equation of state, we can obtain some information about the transition of a black hole. The maximum and minimum of $p$ as functions of $v$ are known as the spinodal curve. This curve separates the (unstable) region of negative compressibility from that of positive compressibility (where the state can be either stable or metastable). This curve is found by solving $T$ in Eq.(\ref{eq:criticality}) and replacing it in Eq.(\ref{PTV}). Thus, we get:  
\begin{equation}
\begin{aligned}
    p=&+\frac{16 q^3-v_c^3}{2 \pi  v_c^5}+\frac{v_c^3-40 q^3}{\pi v_c^5 \left(v_c^3+32 q^3\right)(v_c^3+8 q^3)^{-1}}
   \\
    &-\frac{2^{\frac{2}{\beta }-1} \epsilon  \left(\frac{b}{v_c}\right){}^{2/\beta } \left[(\beta -2) v_c^3-16 (\beta +1) q^3\right]}{\pi  (\beta -2) v_c^5}
    \\
   &+\frac{4^{1/\beta } \epsilon  \left(\frac{b}{v_c}\right){}^{2/\beta } \left[(\beta -2) v_c^3-8 (5 \beta +2) q^3\right]}{\pi  (\beta -2) \beta v_c^5 \left(v_c^3+32 q^3\right)(\beta +1)^{-1}(v_c^3+8 q^3)^{-1}}.
\end{aligned}
\end{equation}

Another interesting piece of information on phase transition is given by Maxwell Construction. This analysis indicates the miscibility gap, which is a two-phase region that separates stable phases. For the van der Walls fluid, this gap separates stable liquid from stable vapor. In the context of black hole thermodynamics, this indicates the region that separates small and large black hole phases. A simple way to determine the miscibility gap graphically is to use the equal area construction of Maxwell.
\par
We can see the spinodal curve and the Maxwell construction for a diagram phase in Fig. (\ref{pv4}), which has phase transition for $T < T_c$ and $T > T_c$ has one phase. In addition, we have six points, of which three points, $a$, $c$, and $e$, are the points with the same pressure $p_0$, which are on the purple dashed line that is the coexistence line between two phases, and the points $b$ and $d$ are the maximum and minimum points, who are on the purple line, which is the spinodal curve. We can also notice that the area $A_1$ has the same value as $A_2$, as was to be expected from Maxwell construction. 

\subsection{Gibbs Free Energy}

In the extended phase space, the mass is interpreted as enthalpy. So the Gibbs free energy can be derived as
\begin{equation}
    G = H - TS = M - TS,
\end{equation}
\noindent where $M$, $T$ and $S$ are given by Eq.(\ref{eq:mass}), (\ref{T_k}) and (\ref{eq:entropy}), respectively. Also using Eq.(\ref{pressure}), we obtain
\begin{equation}
\begin{aligned}
G = &+\frac{r_+ \epsilon  \left(\frac{b}{r_+}\right){}^{2/\beta }}{2 (\beta -2)}-\frac{4}{3} \pi  P r_+^3
\\
&+\frac{3 \beta  \epsilon  \left(\frac{b}{r_+}\right){}^{2/\beta }+3 \beta +8 \pi  \beta  P r_+^2-16 \pi  P r_+^2-6}{\frac{6 (\beta -2) r_+^2}{q^3+r_+^3}}
\\
&+\frac{-3 \beta  r_+ \epsilon  \left(\frac{b}{r_+}\right){}^{2/\beta }-8 \pi  \beta  P r_+^3+16 \pi  P r_+^3-3 \beta  r_++6 r_+}{\frac{12 (\beta -2) \left(q^3+r_+^3\right)}{r_+^3-2 q^3}}.
\end{aligned}
\label{eq:F}
\end{equation}

\begin{figure}[h!]
     \centering
    \includegraphics[width=0.48\textwidth]{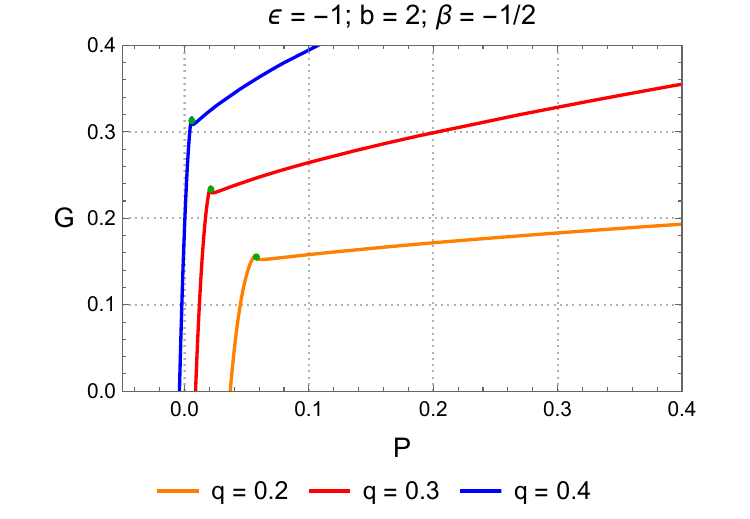}
    \caption{Graphic representation of the $G - P$ behavior for different values of $q$.}
    \label{im5a}
\end{figure}

\begin{figure}[h!]
     \centering
    \includegraphics[width=0.48\textwidth]{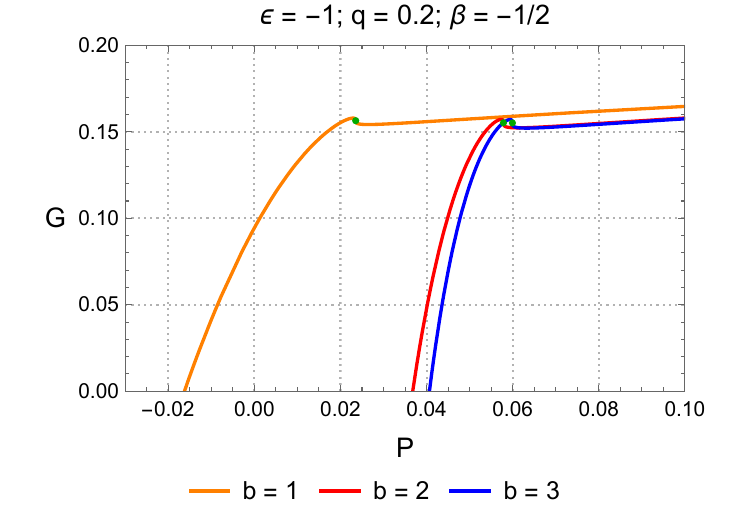}
    \caption{Graphic representation of the $G - P$ behavior of the solution (\ref{f_r}) for different values of $b$.}
    \label{im5b}
\end{figure}

The phase transition can also be verified by the analysis of the Gibbs free energy.
The behavior of Gibbs free energy is depicted in Figs. (\ref{im5a})-(\ref{im5b}) for different values of $q$ and $b$ with temperature critical that can be identified in the Table (\ref{table:critical_val_q})-(\ref{table:critical_val_b}).

\begin{figure}[h!]
     \centering
    \includegraphics[width=0.48\textwidth]{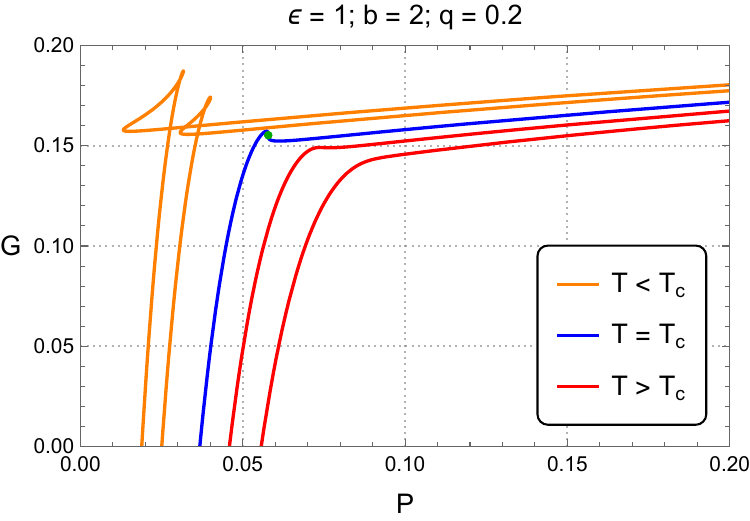}
    \caption{Graphic representation of the $G - P$ behavior of the solution (\ref{f_r}) at a $T<T_c$, $T=T_c$ and $T>T_c$.}
    \label{im6b}
\end{figure}

\begin{figure}[h!]
     \centering
    \includegraphics[width=0.48\textwidth]{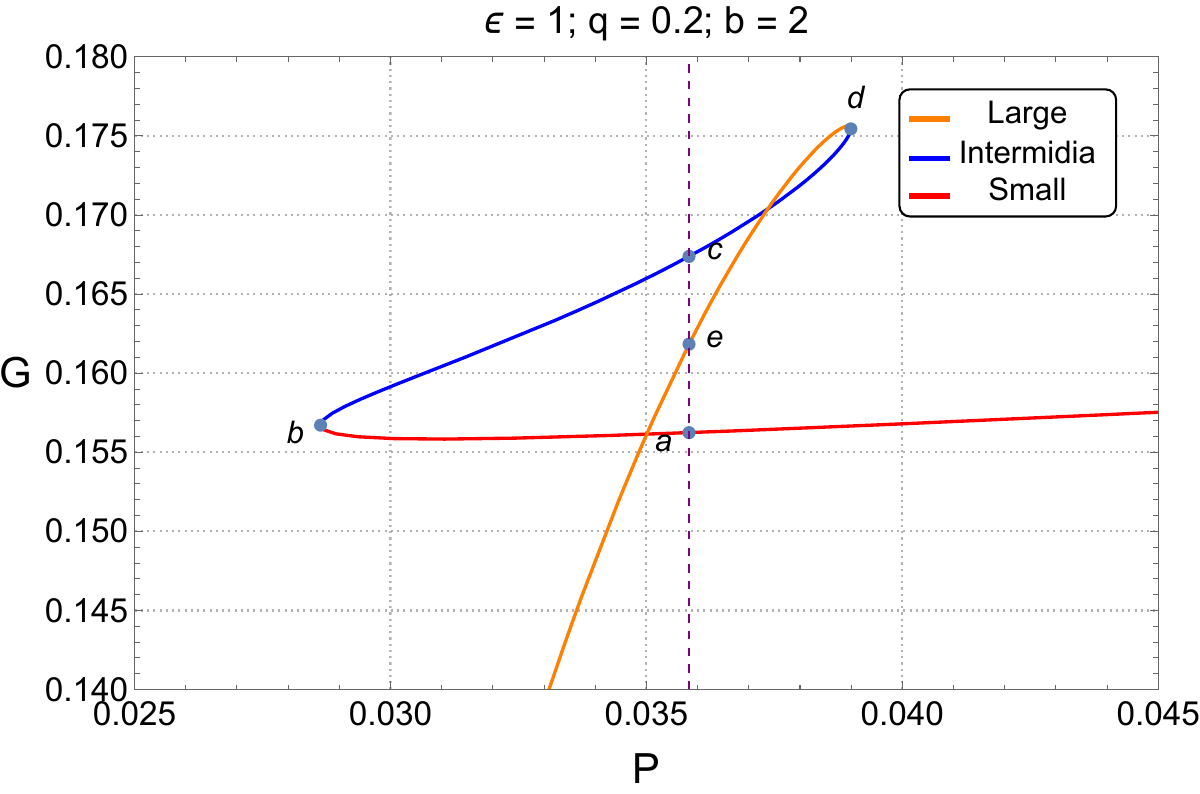}
    \caption{In the graphic, we take a zoom for a curve with $85\%$ of the critical temperature and the orange line is the Large black hole, the blue line is the Intermedia black hole and the red line is the Small black hole.}
    \label{im7b}
\end{figure}

In Fig. (\ref{im6b}) we show the Gibbs free energy for different values of temperature. For $T < T_c$ we have a phase transition; when the temperature increases, the two phases become just one. In the graphic, the blue line is the critical isotherm; the orange lines are the isotherms with temperatures below the critical temperature, and the red lines are the isotherms with temperatures above the critical temperature.

Now in Fig. (\ref{im7b}), we make a zoom in Fig.(\ref{im6b}), in this graphic, we use the same temperature $T = 0.85T_c$ that is used in Fig.(\ref{pv4}) for the phase diagram. Notice that points $a, b, c, d$, and $e$ indicate the regions of phase transition and the unstable and stable transitions. The Gibbs equation gives us information about the phase transition. When we have a temperature below the critical temperature, the curve of $G\times P$ has two tips that can separate the transition phase. In Fig. (\ref{im7b}), it is possible to note this behavior. Notice that, in the red line, we have the small black holes; in the blue line, we have a coexistence between the small and large black holes, which is indicated as an intermediate black hole; and finally, in the orange line, we have the large black hole. Additionally, the points $a, c$, and $e$ have the same pressure, $P_0$, because these points are on the same line of coexistence.

\section{Concluding remarks}\label{sec:concl}

In this article, we have obtained and analyzed a class of solutions that generalizes the original Hayward black hole solution. This generalization is performed by the addition of the cosmological constant and a fluid of strings, which surrounds the Hayward black hole. We have found the metrics that represent this class of spacetimes, and analyzed their regularity or singularity by studying the behavior of the Kretschmann scalar and the geodesics. An interesting and surprising result is that the predictions arising from the analysis of the Kretschmann scalar are confirmed by the analysis of the geodesics.

We also find that the solutions are regular only for values of $-1\leq\beta<0$. In particular, for $\beta = 2$, which can be interpreted as a solution for perfect fluid dark matter, the Kretschmann scalar diverges at the origin, and therefore, it is not regular. On the other hand, if we examine from the point of view of the geodesic, we conclude that it is incomplete, and therefore, by the result arising from the analysis of the Kretschmann scalar.

We also investigated the black hole thermodynamics for $\beta = -1/2$, which is in the interval that Hayward black hole regularity is preserved. In this context, we obtained two temperatures: one is $T_H$ which is found by the first law of thermodynamics without modifications, and the other is $T_\kappa$ by the surface gravity. 

It is worth pointing out that these quantities are not equal, and thus, we have to do modifications in the first law by introducing the factor $W(r_+,q)$, to connect these two temperatures. The same factor that was used to connect the temperatures, was used to connect other thermodynamic quantities, such as the heat capacity. 
 
 Taking into account the heat capacity, we see that there are discontinuities in this quantity, which means that phase transition occurs. Additionally,
 the heat capacity as a function of the entropy assumes negative and positive values, thus indicating that there are regions where the solution is thermodynamically stable $(C_P>0)$ and other regions in which it is unstable $(C_P<0)$. Another point is the analysis of the equation of state, which behaves like a van der Walls fluid. The function pressure by specific volume is found, and its behavior is described for different values of the parameters $q$ and $b$. Together with the equation of state, we obtained the critical values, which are indicated in Tables (\ref{table:critical_val_q}) and (\ref{table:critical_val_b}). We also construct the phase diagram $P\times v$ for different values of temperature with Maxwell's construction and the spinodal curve.

Finally, we analyze the Gibbs free energy and show that the solution has two thermodynamically stable states, namely small and large black holes, and a third phase transition, which is called the intermediate black hole, which is thermodynamically unstable.

The results obtained show us how the regularity of the system under consideration is affected by the presence of the fluid of strings. Additionally, the study of black hole thermodynamics describes how this object behaves. In future works, we hope to explore some topics that are missing in this work, such as the analysis of shadows, lensing, and the construction of the rotating solutions.

\section*{Acknowledgements}
V.B. Bezerra is partially
supported by CNPq-Brazil ( Conselho Nacional de Desenvolvimento Científico e Tecnológico) through Research Project No. 307211/2020-7.

F. F. Nascimento and J. M. Toledo acknowledge Departamento de Fisica, Universidade Federal da Paraíba, for hospitality.


\bibliographystyle{spphys}       
\bibliography{refs}   

\end{document}